\documentclass[11pt, a4paper,english]{article}
\usepackage{graphicx} 
\usepackage[english]{babel}
\usepackage[utf8]{inputenc}
\usepackage[T1]{fontenc}
\usepackage{lmodern}
\usepackage{caption,subcaption}
\usepackage{geometry}
\usepackage{hyperref}
\usepackage{tabularx}
\usepackage{authblk}
\usepackage{overpic,color}
\usepackage{amsmath}
\usepackage{amsthm}
\usepackage{amssymb}
\usepackage{bm}	

\usepackage[doi=false,isbn=false,url=false,eprint=false,maxnames=50,date=year,clearlang,giveninits=true]{biblatex}

\usepackage{csquotes}

\numberwithin{equation}{section}

\renewcommand{\vector}[1]{\boldsymbol{#1}}
\newcommand{\Fr}{\textnormal{Fr}}
\newcommand{\He}{\textnormal{S}}

\newcommand{\dt}{\partial_t}
\newcommand{\dx}{\partial_x}

\title{A Splitting Scheme for Dispersive Shallow Moment Equations}
\author[1]{Ullika Scholz}
\author[1,2]{Robin Paar}
\author[1]{Manuel Torrilhon}

\affil[1]{Chair of Applied and Computational Mathematics}
\affil[2]{Research Training Group EDDy}

\date{\today}

\begin{document}

\maketitle

\begin{abstract}
    The well-known Shallow Water Equations (SWE) are used for modeling incompressible free-surface flows whenever the shallowness allows for a vertical-averaging; i.e., vertical effects are negligible in comparison to horizontal ones. But vertical averaging comes with the price of losing information along the vertical axis. 
    Moment models for shallow flow contain information on the vertical velocity and pressure profile despite being dimensionally reduced.
    A class of these models incorporating a non-hydrostatic pressure have been introduced before as Dispersive Shallow Moment Models (DSM). However, no method for solving the non-stationary equations has been presented yet, mainly because it was unclear how to compute the pressure equation in the form of the divergence-free constraint.
    We rewrite the pressure equations of the DSM models in the form of a Poisson-like problem to enable their solution with a projection-type splitting scheme. For the linear equations, we present the calculations for the generalized model and discuss the non-linear case. We state the first two linear models and the corresponding nonlinear counterparts. Finally, we introduce a hybrid Finite-Volume Finite-Difference method and discuss the non-stationary numerical results for an experiment with periodic boundary and uneven bottom topography.
\end{abstract}

\section{Introduction}\label{sec:introduction}

Free-surface flows occur in many different contexts in nature and industry. Therefore, their efficient computation is an important problem. Numerous models for incompressible free-surface flows are available, many of which are tailored to specific situations of applications. 

A particularly simple and well-known model is the classical Shallow Water Equations (SWE). From a mathematical perspective, it is a hyperbolic system of PDEs derived from the Euler equations via vertical integration; see e.g. \cite{Lannes_Water-Waves-Problem} for a rigorous mathematical derivation of the SWE and other reduced models. A prerequisite for successfully modeling the flow using the vertically integrated SWE model is that the velocity profile of the horizontal flow is nearly constant. 

The SWE model is accurate for shallow flows where the ratio between the fluid height and the vertical length scale is small. If this situation does not apply, two things happen: First, the averaging of the horizontal velocity is no longer justified. One possible solution to this problem is to switch to the -- still dimensionally reduced but slightly more complex -- so-called Shallow Water Moment Equations (SWME) \cite{KowalskiTorrilhon_shallowmomentapprox}. They rely on a polynomial expansion in vertical direction, encoding information on the vertical profile in additional flow variables, also called moments. Since the introduction of the SWME in 2019, these have been further developed in many ways. Some noticeable examples are hyperbolic regularization \cite{KoellermeierRominger2020,Bauerle2025}, bedload transport~\cite{DiazCastroKoellermeierbedload2021}, the combination with multilayer models~\cite{GarresDiaz2023}, the change of the basis functions in the expansion to Chebyshev polynomials or splines~\cite{Steldermann2024,ScholzKoellermeier2025} as well as the integration into larger computational frameworks~\cite{Zhou2025,Steldermann2026}.
On the other hand, as shallowness decreases, the influence of dispersive effects increases, but frequency dispersion is not modeled by the SWE. This means in practice that not only the shape of waves is poorly captured by the SWE, but the speed of wave packages is generally overestimated. This problem can be remedied in mathematical modeling by incorporating a non-hydrostatic pressure in the PDE system. One well-known system in this context are the Green-Naghdi equations \cite{GreenNaghdi1976, FernandezNieto2018}. See \cite{Escalante2020} for more examples of dispersive models and their first-order hyperbolic formulations.

The Dispersive Shallow Moment Equations (DSM) form a hierarchy of reduced models that account for both, complex profile shapes and non-hydrostatic effects. A derivation of the DSM systems has already been carried out in \cite{ScholzKowalskiTorrilhon_dispersionshallowmoment}. There, an analysis of the dispersion relation and first numerical experiments treating stationary flow are provided. In this paper, the focus is on results for nonstationary flows. To this end, the derivation is formalized even more strictly. However, embedding the DSM models into the framework of \cite{Lannes_Water-Waves-Problem} is bound to fail, as the DSM models are generally not rotation-free: Even for the simplest initial conditions as a linear velocity profile the rotation-free condition is not satisfied.

Following the formalized derivation, a suitable numerical method is established for the nonstationary DSM equations. Due to the pressure equation, which is present in the form of a constraint, typical finite volume schemes cannot be applied directly. One way to circumvent the issue is a hyperbolic reformulation of the problem; see \cite{Favrie2017,Escalante2019,Bassi2020,Busto2021}. This paper follows a different path and employs a splitting scheme instead, where in each timestep two decoupled equations are solved. This idea dates back to the first projection methods for incompressible Navier-Stokes or Euler from the 1960s. The necessary reformulation of the constraint into a Poisson problem was first demonstrated by Harlow and Welch \cite{Harlow1965}. According to \cite{Guermond2006}, projection methods can be roughly divided into pressure-correction and velocity-correction schemes with a third type: the closely related so-called splitting schemes, as introduced by the same authors in \cite{Guermond2003}. These schemes do not involve a correction, but rather determine velocity and pressure in two consecutive steps and do not belong to the group of projection schemes in the strict sense, so the corresponding methods were named projection-type. While projection(-type) methods are suitable also for reduced models like the Green-Naghdi equations (see e.g. \cite{Bonneton2011, Parisot2019, Noelle2022}), the hierarchy named R-GN equations in \cite{Zhang2013,Panda2014} and that we refer to as Zhang-Legendre systems, or the DSM hierarchy as shown in \cite{Escalante2024}, not in all cases the incompressibility constraint will transform into a true Poisson equation. A previous mapping of the reference system including the incompressibility constraint allows only the derivation of a Poisson-like equation following the terminology of \cite{Escalante2024} for the pressure which also explicitly depends on the unknown variable and its first derivative. The Poisson or Poisson-like problem is often solved directly using discrete differential operators \cite{Johnston2004,Escalante2024, Bonneton2011} but there are some cases where the problem is solved in the weak form, e.g. with a Galerkin method \cite{Guermond2003}. In contrast to \cite{Escalante2024} where the authors also present dynamic results obtained from dispersive models based on the shallow water moment models the experiments/equations in this paper are given in a more general setting. The DSM models are derived for an arbitrary number of polynomial basis functions in case of the linear model as well as the nonlinear model.

This paper is organized as follows: In section 2 we derive a mapped reference system from the incompressible Euler equations using a notation that incorporates the Jacobian of the mapping. We further show how the reformulation of the constraint is done by taking the constraint of the linear reference system as an example. In section 3 the dimensional reduction takes place. Here, a more systematic notation is used compared to the previous DSM paper \cite{ScholzKowalskiTorrilhon_dispersionshallowmoment}. At the end of this section the generic moment system is presented and the reformulation of the pressure constraint is executed on the generic moment system.
In the subsequent section 4, specific linear and nonlinear DSM systems with their reformulated pressure constraints are presented.
The last section 5 briefly explains the projection-type method and discusses the numerical results.

\section{Reference System}\label{sec:reference_system}

We will first summarize the key points in the derivation of the DSM model starting from the incompressible Free-Surface Euler equations and follow the steps in~\cite{ScholzKowalskiTorrilhon_dispersionshallowmoment}, but formalize the notation more strictly.\\
We consider an incompressible and inviscid fluid in a three dimensional and time-dependent domain 
\begin{equation}
    \label{eq:domain}
    \Omega_t = \{ (x,y,z) \in \mathbb{R}^3 |\, h_b(x,y) < z < h_s(t,x,y) \}
\end{equation}
as reference surface-flow. Here \(h_b(x,y)\) and \(h_s(t,x,y)\) are parameterizations of the basal topography and the upper free-surface, respectively. Note that we assume the bottom to be time independent, an assumption that has to be relaxed for instance if there is a mass intake. We obtain the standard coordinate system by assuming the gravitational acceleration to be of the form \(\boldsymbol{g} = g(0,0,-1)^T\).
We neglect viscous forces \(\boldsymbol{\sigma}\) and assume the density \(\rho\) to be constant such that we get a divergence constraint on the velocity \(\boldsymbol{u} = (u,v,w)^T\). All three components of the velocity depend on \(t,x,y,z\). Due to the constant density we are also able to write the pressure \(p(t,x,y,z)\) in a density-scaled form. All of these assumptions correspond to the case of liquid water.
Hence, the incompressible Free-Surface Euler equations in this case are given by
\begin{subequations}\label{eq:free-surface-Euler_ref}
    \begin{align}
        \label{eq:free-surface-Euler_ref_evolution}
        \partial_t\vector{u} + \left(\vector{u} \cdot \nabla\right)\vector{u} &= -\nabla p + \vector{g},\\
        \label{eq:free-surface-Euler_ref_divergence}
        \nabla \cdot \vector{u} &= 0.
    \end{align}
\end{subequations}
with the standard nabla operator \(\nabla = (\partial_x,\partial_y,\partial_z)^{\rm T}\).
To write down the boundary conditions, we define the normal vectors at the surface and bottom by
\begin{subequations}\label{eq:domain-unit-normal-vectors}
    \begin{align}
    \label{eq:unit_normal_vector_surface}
    	\vector{n}_s(t,x,y) &= \left(\partial_xh_s(t,x,y),\partial_yh_s(t,x,y),-1\right)^{\rm T},\\
        \label{eq:unit_normal_vector_bottom}
    	 \vector{n}_b(x,y) &= \left(\partial_xh_b(x,y), \partial_yh_b(x,y), -1\right)^{\rm T}.
    \end{align}
\end{subequations}
With this notation, we can now state our kinematic boundary conditions in the following form
\begin{subequations}\label{eq:free-surface-Euler-boundary}
    \begin{equation}
        \label{eq:free-surface-Euler_ref_surface}
        \partial_t h_s + \vector{u}_{|_{z=h_s}} \cdot \vector{n}_s = 0
    \end{equation}
    and
    \begin{equation}
        \label{eq:free-srufac-Euler_ref_bottom}
        \vector{u}_{|_{z=h_b}} \cdot \vector{n}_b = 0.
    \end{equation}
\end{subequations}
The equations~\eqref{eq:free-surface-Euler_ref} and~\eqref{eq:free-surface-Euler-boundary} form our reference system, which we will use as a starting point for all further studies.

\subsection{Mapping of the Reference System}\label{subsec:mapped_ref_sys}

\begin{figure}
   \centering
   \includegraphics[width=0.9\linewidth]{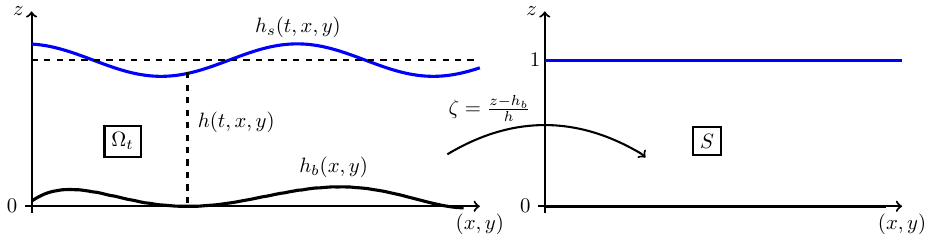}
    \caption{Mapping of the domain \(\Omega_t\) onto the flat strip \(S\).}
    \label{fig:mapping_domain}
\end{figure}

We want to map our reference system to a scaled and normalized vertical coordinate referred to as \(\zeta\). In the literature this new coordinates are often referred to as sigma-coordinates after~\cite{Phillips1957}. We define the height of the fluid as the difference of free-surface and basal topography, hence
\begin{equation}
    \label{eq:def_height}
	h(t,x,y) := h_s(t,x,y) - h_b(x,y).
\end{equation}
This quantity is of central interest in any type of shallow flow models.
This definition gives us an convenient way to define our new coordinates
\begin{equation}\label{eq:def_sigma_coord}
	\zeta = \frac{z - h_b(x,y)}{h(t,x,y)}.
\end{equation}
If we define the flat strip through
\begin{equation}
    \label{eq:flat_strip_domain}
    S = \left\{ (x,y,\zeta)\in\mathbb{R}^3 | \, 0 < \zeta < 1\right\},
\end{equation}
which is independent of the time, we can formalize this mapping with a function
\begin{equation}
    \label{eq:mapping_function}
    \Sigma: \mathbb{R}\times \Omega_t \rightarrow \mathbb{R}\times S , (t,x,y,z) \mapsto \Sigma(t,x,y,z) = \left(t,x,y, \frac{z - h_b}{h}\right) = (t,x,y,\zeta).
\end{equation}
This kind of mapping is illustrated in figure~\ref{fig:mapping_domain}.
With the mapping~\(\Sigma\) we have for an arbitrary function~\(\psi = \psi(t,x,y,z)\)
\begin{equation}
\label{eq:relation_functions}
	\psi(t,x,y,z) = \left(\tilde{\psi}\circ\Sigma\right)(t,x,y,z) = \tilde{\psi}\left(t,x,y,\frac{z - h_b}{h}\right) = \tilde{\psi}(t,x,y,\zeta). 
\end{equation}
Now, we are able to express derivatives of the original function \(\psi\) in terms of the mapped function~\(\tilde{\psi}\). Before we do so, we first define the differential operators \(D=(\partial_t,\partial_x,\partial_y,\partial_z)^{\rm T}\), resp., \(\tilde{D}=(\partial_t,\partial_x,\partial_y,\partial_{\zeta})^{\rm T}\), giving us the relation
\begin{equation}
\label{eq:relation_gradients}
    D\psi = D\left(\tilde{\psi}\circ\Sigma\right) = \left(J_{\tilde{\psi}\circ\Sigma}\right)^{\rm T} = \left(J_{\tilde{\psi}} J_{\Sigma}\right)^{\rm T} =\left(J_{\Sigma}\right)^{\rm T} \left(J_{\tilde{\psi}}\right)^{\rm T} = \left(J_{\Sigma}\right)^{\rm T}\tilde{D}\tilde{\psi}.
\end{equation}
We used the multidimensional chain rule together with the Jacobian matrices of both mappings \(\tilde{\psi}\) and \(\Sigma\). The Jacobians of \(\Sigma\) and \(\psi = \tilde{\psi}\circ\Sigma\) are defined in terms of \((t,x,y,z)\) and the Jacobian of \(\tilde{\psi}\) in terms of \((t,x,y,\zeta)\). Remark, from the definition of the Jacobian the relation \(D\psi = \left(J_{\psi}\right)^{\rm T}\) for a scalar valued function \(\psi\) holds. So knowing the Jacobian of \(\Sigma\) we can directly map the derivatives following equation~\eqref{eq:relation_gradients}. We calculate with \(\Sigma\) given by~\eqref{eq:mapping_function}
\begin{equation}
    \label{eq:jacobian_sigma}
    J_{\Sigma} = \begin{pmatrix}
        1 & 0 & 0 & 0 \\ 0 & 1 & 0 & 0 \\ 0 & 0 & 1 & 0 \\ \partial_t\zeta & \partial_x \zeta & \partial_y \zeta & \partial_z \zeta
    \end{pmatrix} = \begin{pmatrix}
        1 & 0 & 0 & 0 \\ 0 & 1 & 0 & 0 \\ 0 & 0 & 1& 0 \\ -\frac{1}{h}\partial_t(\zeta h + h_b) & -\frac{1}{h}\partial_x(\zeta h + h_b) & -\frac{1}{h}\partial_y(\zeta h + h_b) & \frac{1}{h}
    \end{pmatrix},
\end{equation}
where we used
\begin{equation}
\label{eq:rel_deri_sigma_coord}
	\partial_s\zeta = \partial_s\frac{z - h_b}{h}
    =-\frac{1}{h}\partial_s(\zeta h + h_b)\quad\text{and}\quad \partial_z\zeta = \partial_z\frac{z - h_b}{h} = \frac{1}{h}
\end{equation}
for \(s\in \{t,x,y\}\).
For clarity, we decompose the gradient in a time-dependent one-dimensional and spatial three-dimensional part, i.e., we have 
\begin{equation}
    \label{eq:grad_decomposition}
    \left(J_{\Sigma}\right)^{\rm T}\tilde{D} = \left(\begin{array}{c}
         \left((J_{\Sigma})^{\rm T}\tilde{D}\right)_t  \\
        \left((J_{\Sigma})^{\rm T}\tilde{D}\right)_x
    \end{array}\right)\quad\text{and}\quad \tilde{D}= \left(\begin{array}{c}
         \partial_t \\
          \tilde{\nabla}
    \end{array}\right), \quad\text{with }\tilde{\nabla} = (\partial_x,\partial_y,\partial_{\zeta})^{\rm T}
\end{equation}
and for practical reasons, we also define \(\tilde{\nabla}_x = (\partial_x,\partial_y,0)^{\rm T}\).
Hence, we can write the mapped free-surface Euler equations in a similar way to~\eqref{eq:free-surface-Euler_ref}
\begin{subequations}\label{eq:free-surfcae-Euler-ref-mapped}
    \begin{align}
        \label{eq:free_surface_Euler_ref_evolution_mapped}
        \left((J_{\Sigma})^{\rm T}\tilde{D}\right)_t\tilde{\boldsymbol{u}} + \left(\tilde{\boldsymbol{u}}\cdot \left((J_{\Sigma})^{\rm T}\tilde{D}\right)_x\right)\tilde{\boldsymbol{u}} &= -\left((J_{\Sigma})^{\rm T}\tilde{D}\right)_x\tilde{p} + \boldsymbol{g},\\
        \label{eq:free_surface_Euler_ref_divergence_mapped}
        \left((J_{\Sigma})^{\rm T}\tilde{\nabla}\right)_x\cdot \tilde{\boldsymbol{u}} &= 0.
    \end{align}
\end{subequations}
The mapped boundary conditions are given by
\begin{subequations}\label{eq:mapped-ref-sys-boundary}
    \begin{align}
        \label{eq:mapped_ref_sys_surface}
        \partial_th_s + \tilde{\boldsymbol{u}}_{|_{\zeta=1}} \cdot \boldsymbol{n}_s &= 0,\\
        \label{eq:mapped_ref_sys_bottom}
        \tilde{\boldsymbol{u}}_{|_{\zeta=0}} \cdot \boldsymbol{n}_b &=0.
    \end{align}
\end{subequations}
The system~\eqref{eq:free-surfcae-Euler-ref-mapped} and~\eqref{eq:mapped-ref-sys-boundary} can be written in less abstract and more accessible way analog to~\cite{ScholzKowalskiTorrilhon_dispersionshallowmoment}. We drop the tilde at the variables for a better readability. More information regarding this reformulation process can be found in the appendix~\ref{appx:subsec:equiv_ref_sys}. The reformulation that uses the definition of the mapped derivatives in a more direct way, is given by
\begin{subequations}\label{eq:free-surface-Euler-comp}
\begin{align}
    \label{eq:free_surface_Euler_height}
    &\partial_th + \partial_x(hu_m) + \partial_y(hv_m) = 0,\\
    \label{eq:free_surface_Euler_divergence}
	&\partial_x(h u) +\partial_y(hv) +\partial_{\zeta} w -\partial_{\zeta}(\partial_x(\zeta h + h_b)u)-\partial_{\zeta}(\partial_y(\zeta h + h_b)v)=0\\
    \label{eq:free_surface_Euler_evolution}
	&\partial_t(h\vector{u}) + \partial_x(hu\vector{u}) + \partial_y(hv\vector{u}) +\partial_{\zeta}(h\omega\vector{u})
	=-\begin{pmatrix}
		\partial_x(hp)-\partial_{\zeta}(\partial_x(\zeta h + h_b)p)\\\partial_y(hp)-\partial_{\zeta}(\partial_y(\zeta h + h_b)p)\\ \partial_{\zeta}p
	\end{pmatrix}+h\vector{g}.
\end{align}
\end{subequations}
In the equations appears the vertical coupling operator \(\omega\) that, following the notation from~\cite{KowalskiTorrilhon_shallowmomentapprox}, is given by
\begin{equation}
	\label{eq:vertical_coupling_op}
	h\omega\left[h,u,v\right] = -\partial_x\left(h\int_{0}^{\zeta}(u - u_m)\,\mathrm{d}\hat{\zeta}\right) - \partial_y\left(h\int_{0}^{\zeta}(v - v_m)\,\mathrm{d}\hat{\zeta}\right),
\end{equation}
where we introduced the vertically averaged values of the velocity components \(u\) and \(v\)
\begin{equation}
\label{eq:vert_averaged_velo}
u_m(t,x,y) = \int_{0}^{1} u(t,x,y,\zeta)\,\mathrm{d}\zeta\qquad \text{and}\qquad v_m(t,x,y) = \int_{0}^{1}v(t,x,y,\zeta)\,\mathrm{d}\zeta.
\end{equation}
There is an interesting relation for \(\omega\)
\begin{equation}
	\label{eq:relation_vert_coupling}
	\partial_x(hu) + \partial_y(hv) + \partial_{\zeta}(h\omega) = \partial_x(hu_m) + \partial_y(hv_m),
\end{equation}
that allows us to replace the height balance~\eqref{eq:free_surface_Euler_height} by 
\begin{equation}
	\label{eq:free_surface_Euler_height_vert_op}
	\partial_th + \partial_x(hu) + \partial_y(hv) + \partial_{\zeta}(h\omega)= 0.
\end{equation}
We will use both formulations in this paper depending on the situation, where either the abstract one or the formulation by components is more suitable.

\subsection{Reference System with Non-Hydrostatic Pressure}\label{subsec:non-hydro_ref_sys}

So far, we made no assumptions on the form of the pressure \(p\). Very often, the Shallow Water Equations are only used with a hydrostatic pressure and the same holds true for the Shallow Water Moment Equations~\cite{KowalskiTorrilhon_shallowmomentapprox}.
In contrast, we assume that the pressure does consists of the hydrostatic part and a second non-hydrostatic part or a so-called deviation. Hence, we can write the pressure in the form
\begin{equation}
\label{eq:def_pressure}
	p(t,x,y,z)=p_{\text{hyd}} + q = (h_s(t,x,y)-z) g + q(t,x,y,z),
\end{equation}
notice, that we have already scaled the pressure by the density \(\rho\).
For our purpose we map this relation to the \(\zeta\)-coordinates introduced before and get
\begin{equation}
	\label{eq:def_pressure_mapped}
	\tilde{p}(t,x,y,\zeta) = (h_s - \zeta h - h_b)g +\tilde{q} = (1-\zeta)hg + \tilde{q}(t,x,y,\zeta). 
\end{equation}
Again, we drop the tilde for readability and substitute this relation for the general pressure term in the mapped Euler equations~\eqref{eq:free-surface-Euler-comp}. The pressure term in the Euler equations now reads (in component form):
\begin{align*}
	-h\left((J_{\Sigma})^{\rm T}\tilde{D}\right)_xp+h\boldsymbol{g}=
	-\frac{g}{2}\tilde{\nabla}_x\left(h^2\right) - hg\tilde{\nabla}_xh_b - \left((J_{\Sigma})^{\rm T}\tilde{D}\right)_xq.
\end{align*}
Two boundary conditions hold: At the bottom, where \(\zeta=0\) the kinematic boundary condition yields
\begin{equation}
\label{eq:bottom_boundary}
	\boldsymbol{u}_{|_{\zeta =0}}\cdot \boldsymbol{n}_b =0 \quad\Leftrightarrow\quad w_{|_{\zeta=0}}=u_{|_{\zeta =0}}\partial_xh_b+v_{|_{\zeta=0}}\partial_yh_b.
\end{equation}
Further, the pressure is scaled in such way that it is zero at the surface. This corresponds to assuming a stress-free state at the free-surface and using that there is no hydrostatic pressure at the surface. From the definition of the pressure this directly yields
\begin{equation}
\label{eq:surface_boundary}
0=p_{|_{\zeta = 1}} = q_{|_{\zeta = 1}} \quad \Leftrightarrow\quad q_{|_{\zeta=1}} = 0,
\end{equation}
which gives us a second boundary condition. This allows us to state our full reference system in its final form, that will be used throughout this paper. We write it down in the same manner as in~\cite{ScholzKowalskiTorrilhon_dispersionshallowmoment}, but keep our notation for the velocity evolution in vectorial form:
\begin{subequations}\label{eq:ref-sys}
\begin{align}
    \label{eq:ref_sys_height}
	&\partial_th + \partial_x(hu) + \partial_y(hv) + \partial_{\zeta}(h\omega) =0,\\
	\label{eq:ref_sys_velocity}
	&\partial_t(h\vector{u}) + \partial_x(hu\vector{u}) + \partial_y(hv\vector{u}) +\partial_{\zeta}(h\omega\vector{u})+\frac{g}{2}\tilde{\nabla}_x\left(h^2\right) + \left((J_{\Sigma})^{\rm T}\tilde{D}\right)_xq = -hg\tilde{\nabla}_xh_b,
\end{align}
with the mapped mass balance resp. the divergence-free condition
\begin{equation}
    \label{eq:ref_sys_constraint}
    \partial_{\zeta} w + \partial_x(h u) +\partial_y(hv) -\partial_{\zeta}(\partial_x(\zeta h +h_b)u)-\partial_{\zeta}(\partial_y(\zeta h + h_b)v)=0.
\end{equation}
\end{subequations}
Along with the system we have the two boundary conditions
\begin{subequations}\label{eq:ref-sys-boundary}
    \begin{equation}
        \label{eq:ref_sys_pressure_boundary}
        q_{|_{\zeta=1}} = 0
    \end{equation}
    and
    \begin{equation}
        \label{eq:ref_sys_velo_boundary}
        w_{|_{\zeta=0}} = u_{|_{\zeta = 0}} \partial_xh_b + v_{|_{\zeta = 0}}\partial_yh_b,
    \end{equation}
\end{subequations}
which describe the value of the non-hydrostatic pressure at the free-surface as well as the value of the vertical velocity component at the bottom. Note that for numerical simulations additional boundary conditions at the sides of the domain that may depend on inflow and outflow conditions are necessary. The system~\eqref{eq:ref-sys} includes non-hydrostatic pressure and is vertically resolved since the variables depend on the vertical direction \(\zeta\).

\subsection{Linear Reference System}\label{subsec:lin_ref_sys}

Before we consider the general case and derive the moment system according to the reference system~\eqref{eq:ref-sys}, we will study a simplified version of the equations obtained through linearization. The corresponding linear reference system can be obtained from the nonlinear system by performing a small perturbation ansatz as explained in \cite{ScholzKowalskiTorrilhon_dispersionshallowmoment}. This corresponds to replacing
\begin{subequations}\label{eq:perturb-ansatz}
\begin{align}
\label{eq:perturb_ansatz_height}
h(t,x,y)&\to \varepsilon \hat{h}(t,x,y)+h_0,\\
\label{eq:perturb_ansatz_velo}
\boldsymbol{u}(t,x,y,\zeta)&\to \varepsilon \boldsymbol{\hat{u}}(t,x,y,\zeta)+\boldsymbol{u}_0,\\
\label{eq:perturb_ansatz_deviation}
q(t,x,y,\zeta)&\to \varepsilon \hat{q}(t,x,y,\zeta),\\
\label{eq:perturb_ansatz_topo}
h_b(x,y)&\to \varepsilon \hat{h_b}(x,y),
\end{align}
\end{subequations}
in~\eqref{eq:ref-sys} and neglecting terms of order $\varepsilon^2$ or higher. The linearized system can be used to accurately describe flows close to a constant ground state. The introduced vector for the velocity ground state is of the form \(\boldsymbol{u}_0 = (u_0,v_0,0)^T\), because the only physically reasonable ground state for the vertical velocity is the vanishing one. The reason is that we consider a gravitational acceleration that is parallel to the vertical axis and we also have only small variety in the bottom topography, such that there is nothing causing a notable vertical velocity. 

Before the reference system is linearized we consider how linearization affects the Jacobian matrix of the transformation \(\Sigma\). With the small perturbation ansatz~\eqref{eq:perturb-ansatz} we first have for~\(s\in\{t,x,y\}\)
\begin{equation}
\label{eq:lin_deri_sigma}
	-\frac{1}{h}\partial_s (\zeta h + h_b) = -\frac{1}{h_0 +\varepsilon\hat{h}}\partial_s\left(\zeta \left(h_0+\varepsilon\hat{h}\right)+\varepsilon\hat{h}_b\right) =-\frac{h_0-\varepsilon\hat{h}}{h_0^2}\varepsilon\partial_s\left(\zeta\hat{h}+\hat{h}_b\right) = -\frac{\varepsilon}{h_0}\partial_s\left(\zeta\hat{h}+\hat{h}_b\right).
\end{equation}
Using this for the Jacobian of the mapping $\Sigma$ given in~\eqref{eq:jacobian_sigma}, we find a decomposition of the matrix similar to the small perturbation ansatz 
\begin{equation}
	\label{eq:jacobian_decomp}
	J_{\Sigma} = J^{(0)}_{\Sigma} + \frac{\varepsilon}{h_0} J^{(1)}_{\Sigma} = \begin{pmatrix}
		1 & 0 & 0 & 0 \\ 0 & 1 & 0 & 0 \\ 0 & 0 & 1 & 0 \\ 0 & 0 & 0 & \frac{1}{h_0}
	\end{pmatrix} + \frac{\varepsilon}{h_0} \begin{pmatrix}
	0 & 0 & 0 & 0 \\ 0 & 0 & 0 & 0 \\ 0 & 0 & 0 & 0\\ -\partial_t\left(\zeta\hat{h}+\hat{h}_b\right) & - \partial_x\left(\zeta \hat{h} + \hat{h}_b\right) & - \partial_y\left(\zeta \hat{h} + \hat{h}_b\right) & -\frac{\hat{h}}{h_0}
	\end{pmatrix}.
\end{equation}
The small perturbation ansatz~\eqref{eq:perturb-ansatz} and the matrix decomposition~\eqref{eq:jacobian_decomp} are then substituted into the mapped free-surface Euler equations~\eqref{eq:free-surfcae-Euler-ref-mapped} and all terms of order \(\varepsilon^2\) are neglected. Note that the matrix \(J_{\Sigma}\) always appears with a derivative of order \(\varepsilon\). Consequently, the term with \(J_{\Sigma}^{(1)}\) is always of order \(\mathcal{O}(\varepsilon^2)\) and vanishes. The resulting linearized free-surface Euler equations read
\begin{subequations}\label{eq:lin-Euler-eq}
	\begin{align}
		\label{eq:lin_Euler_evol}
		\left(\left(J_{\Sigma}^{(0)}\right)^{\rm T}\tilde{D}\right)_t\boldsymbol{\hat{u}} + \left(\boldsymbol{u}_0\cdot\left(\left(J_{\Sigma}^{(0)}\right)^{\rm T}\tilde{D}\right)_x\right)\boldsymbol{\hat{u}} &= -\left(\left(J_{\Sigma}^{(0)}\right)^{\rm T}\tilde{D}\right)_x\hat{p} +\boldsymbol{g},\\
		\label{eq:lin_Euler_div}
		\left(\left(J_{\Sigma}^{(0)}\right)^{\rm T}\tilde{D}\right)_x\cdot\boldsymbol{\hat{u}} &= 0.
	\end{align}
\end{subequations}
 The definition~\eqref{eq:def_pressure} of the general pressure \(p\) can be linearized very easily. To keep the equations simple we define the gradient \(\hat{\nabla}:=\left(\partial_x,\partial_y,\frac{1}{h_0}\partial_{\zeta}\right)^{\rm T}\) and omit the hat on our linearized variables. In addition, we define the vector of the mean velocity \(\boldsymbol{u}_m = (u_m,v_m,w_m)\), where \(w_m\) is defined in the same way as the other two components following~\eqref{eq:vert_averaged_velo}. Finally, 
\begin{subequations}\label{eq:lin-ref-sys}
	\begin{align}
		\label{eq:lin_ref_sys_height}
		\partial_th + h_0\hat{\nabla}\boldsymbol{u}_m + \boldsymbol{u}_0\cdot\hat{\nabla}h &= 0,\\
		\label{eq:lin_ref_sys_evol}
		\partial_t\boldsymbol{u} + \left(\boldsymbol{u}_0\cdot\hat{\nabla}\right)\boldsymbol{u} + g\hat{\nabla}h + \hat{\nabla}q &= -g\hat{\nabla}h_b,\\
		\label{eq:lin_ref_sys_div}
		\hat{\nabla} \cdot \boldsymbol{u} &= 0,
	\end{align}
	with the two boundary conditions
	\begin{equation}
		\label{eq:lin_ref_sys_surface_boundary}
		q_{|_{\zeta = 1}} =0 
	\end{equation}
	and
	\begin{equation}
		\label{eq:lin_rf_sys_bottom_boundary}
		w_{|_{\zeta = 0}} = \boldsymbol{u}_0 \cdot \hat{\nabla}h_b.
	\end{equation}
\end{subequations}
are the equations that will serve as linearized reference system, written in the manner introduced in section~\eqref{subsec:non-hydro_ref_sys}.

\subsection{Pressure Equation for the Linear Reference System}\label{subsec:pressure_eq_lin_ref_sys}

It is possible to replace the divergence constraint in the linear reference system~\eqref{eq:lin_ref_sys_div} with a Poisson problem for the pressure deviation \(q\). Later for the (linear) dispersive shallow moment systems the steps are the same as for the (linear) reference system and hence, we will show the calculation for the linear reference system at this point. We start taking the divergence \(\hat{\nabla}\cdot\) of the velocity evolution~\eqref{eq:lin_ref_sys_evol}
\begin{equation}\label{eq:lin_ref_pressure_poisson_div}
	\hat{\nabla}\cdot\partial_t\boldsymbol{u} + \hat{\nabla}\cdot\left(\left(\boldsymbol{u}_0\cdot\hat{\nabla}\right)\boldsymbol{u}\right) + g\hat{\nabla}\cdot\hat{\nabla}h + \hat{\nabla}\cdot\hat{\nabla}q = -g\hat{\nabla}\cdot\hat{\nabla}h_b.
\end{equation}
In the next step we use that all derivatives commute and the definition of the Laplacian \(\hat{\Delta}=\hat{\nabla}\cdot\hat{\nabla}\) to achieve
\begin{equation}\label{eq:lin_ref_pressure_poisson_reformulation}
	\partial_t\left(\hat{\nabla}\cdot\boldsymbol{u}\right) + \left(\boldsymbol{u}_0\cdot\hat{\nabla}\right)\left(\hat{\nabla}\cdot\boldsymbol{u}\right) + \left(\hat{\nabla}\boldsymbol{u}_0\right)^T:\left(\hat{\nabla}\boldsymbol{u}\right) + g\hat{\Delta}h + \hat{\Delta}q = - g\hat{\Delta}h_b.
 \end{equation} 
Taking into account that the mapped variable \(\boldsymbol{u}\) is divergence-free for the mapped divergence operator $\hat{\nabla}\cdot$ and  \(\boldsymbol{u}_0\) is constant, we get the Poisson equation
\begin{equation}
	\label{eq:lin_ref_sys_pressure}
	\hat{\Delta}q = -g\hat{\Delta}(h + h_b).
\end{equation}
We have replaced the divergence-free condition with an equation for the pressure deviation that can be used to solve the system using a pressure splitting scheme. More details on this splitting scheme are given in section~\ref{sec:Num_Experiments}.

\section{General Shallow Moment Equations with Non-Hydrostatic Pressure}\label{sec:general_moment_systems}

The starting point for all manipulations are the mapped incompressible free-surface Euler equations from section~\ref{subsec:non-hydro_ref_sys} respective the linearized equations from section~\ref{subsec:lin_ref_sys}.
First, we will derive the Linear Dispersive Shallow Moment Systems in detail using a more strict and formalized notation. We show how to derive an equation for the pressure deviation like in section~\ref{subsec:pressure_eq_lin_ref_sys} for the general case. For the DSM systems a detailed derivation of the equations can be found in the appendix~\ref{appx:subsec:DSM-N} that follows the derivative from~\cite{ScholzKowalskiTorrilhon_dispersionshallowmoment}, but uses a new introduced vectorial notation. Here we state the equations and give a schematic way to derive the pressure equations for the DSM systems.

\subsection{Linear Dispersive Shallow Moment Equations (LDSM)}\label{subsec:LDSM}

The LDSM have a lower dimensionality compared to the original system but keep information on the vertical profiles of the fields. For this propose, we use the standard moment approach, i.e., we assume that all quantities are polynomial in \(\zeta\)-direction, multiply by the different basis functions up to degree \(N\) and integrate vertically. The highest degree of the basis function \(N\) will also be called the level of the moment system \cite{KowalskiTorrilhon_shallowmomentapprox}. 

As basis of the polynomial space we use on the interval \(\left[0,1 \right]\) orthogonal Legendre polynomials \(\phi_j\) that are normed at the bottom, i.e., \(\phi_j(0) = 1\). For these polynomials we have additional properties that are useful for the derivation of the equations. For example, on the other boundary of the domain it holds \(\phi_j(1) = (-1)^j\) and the orthogonality factor is explicitly given by
\begin{equation}
    \label{eq:legendre_orthogonal}
    \int_0^1 \phi_i(\zeta)\phi_j(\zeta)\,\mathrm{d}\zeta = \frac{1}{2i+1}\delta_{ij}.
\end{equation}
In contrast to previous papers we use a vectorial notation consistent with the derivations in section~\ref{sec:reference_system}. For the linear case we will illustrate the whole derivation of the moment system using the new notation and introduce a more accurate way to write the equations down. This also gives us a more convenient way to derive and state the equations for the mean pressure deviation and all the higher pressure moments.
Therefore, recall the notation from the previous section: 
\begin{equation}
\label{eq:notation_moments}
\hat{\nabla}=\left(\partial_x,\partial_y,\frac{1}{h_0}\partial_\zeta\right)^{\rm T},\,\vector{u}=(u,v,w)^{\rm T},\,\vector{u}_0=(u_0,v_0,0)^{\rm T},\,\vector{u}_m = (u_m,v_m,w_m)^{\rm T}.
\end{equation}
We assume the velocity profile to be of the form
\begin{equation}
\label{eq:velocity_moments}
	\vector{u}(t,x,y,\zeta) = \vector{u}_m(t,x,y) + \vector{u}_d(t,x,y,\zeta) = \vector{u}_m(t,x,y) + \sum_{i=1}^{N}\vector{\Lambda}_i(t,x,y)\phi_i(\zeta),
\end{equation}
with the coefficient vector
\begin{equation}
\label{eq:velocity_moments_coeff}
	\vector{\Lambda}_i(t,x,y) = (\alpha_i(t,x,y),\beta_i(t,x,y),\gamma_i(t,x,y))^{\rm T}
\end{equation}
and similar for the pressure deviation
\begin{equation}
\label{eq:pressure_moments}
	q(t,x,y,\zeta) = q_m(t,x,y) + \sum_{i=1}^{N}\kappa_i(t,x,y)\phi_i(\zeta).
\end{equation}
We take the same polynomial ansatz for the velocity profile and the pressure deviation. This way we can use the orthogonality of the Legendre polynomials in full extent and end up with a hierarchical model.
We use the ansatz~\eqref{eq:velocity_moments} and~\eqref{eq:pressure_moments} in our linearized reference system~\eqref{eq:lin-ref-sys} and integrate the equations in the vertical direction using the Galerkin projections
\begin{equation*}
    \int_0^1 \bullet\,\phi_j\,\mathrm{d}\zeta,\qquad\text{for }j=0,\dots,N.
\end{equation*}
Remark, for the case \(j=0\) we just have \(\phi_0\equiv 1\), that means the projection corresponds to the vertical averaging procedure used for the classical shallow water equations.
Before we consider the dimensional reduction of the linear reference system in more detail, we first keep our attention on the boundary conditions.
Inserting our ansatz into the boundary conditions we can replace one of the coefficients of the pressure deviation and vertical velocity component by a linear combination of the other
\begin{subequations}\label{eq:moment-boundary}
\begin{align}
\label{eq:moment_boundary_surface}
	&q_{|_{\zeta=1}} = q_m + \sum_{j=1}^{N}\kappa_j\phi_j(1) = q_m + \sum_{j=1}^{N}\kappa_j(-1)^j = 0\quad \Leftrightarrow\quad q_m = \sum_{j=1}^{N}(-1)^{j+1}\kappa_j,\\
    \label{eq:moment_boundary_bottom}
	&w_{|_{\zeta =0}} = w_m + \sum_{j=1}^{N}\gamma_j\phi(0) = w_m+\sum_{j=1}^{N}\gamma_j = \vector{u}_0\cdot\hat{\nabla}h_b\quad\Leftrightarrow\quad w_m = \boldsymbol{u}_0\cdot\hat{\nabla}h_b - \sum_{j=1}^N\gamma_j.
\end{align}
\end{subequations}
This means for the reduced system to be overdetermined by two equations if the level \(N\) is greater than zero. We avoid this problem by using the boundary conditions only in a weak sense and not enforcing them strongly, see 
\cite{KowalskiTorrilhon_shallowmomentapprox} for a detailed discussion. More details will be given later in the derivation of the LDSM model.
To keep the notation simple, we introduce the vectors of variables \(\boldsymbol{\gamma} = (\alpha_1,\dots,\alpha_N)^{\rm T}\) and \(\boldsymbol{\kappa}=(\kappa_1,\dots,\kappa_N)^{\rm T}\). In a similar way we can also define \(\boldsymbol{\alpha} = (\alpha_1,\dots,\alpha_N)^{\rm T}\) that we will need later.

Since the height balance~\eqref{eq:lin_ref_sys_height} does not depend on the vertical variable it is not affected by the dimensional reduction. Continuing with the velocity evolution~\eqref{eq:lin_ref_sys_evol}, multiplying with a test function \(\phi_i\) and integrating we get the reduced equations
\begin{subequations}\label{eq:moments-evol}
\begin{equation}
\label{eq:moments_evol_zero}
		\partial_t\vector{u}_m + (\vector{u}_0\cdot\hat{\nabla})\vector{u}_m +g\hat{\nabla}h +\hat{\nabla}q_m = -g\hat{\nabla}h_b + A_0^N(q_m,\boldsymbol{\kappa})\vector{e}_{\zeta}
        \end{equation}
        for \(i=0\) and
        \begin{equation}
        \label{eq:moments_evol_higher}
		\partial_t\vector{\Lambda}_i +(\vector{u}_0\cdot\hat{\nabla})\vector{\Lambda}_i + \hat{\nabla}\kappa_i = (2i+1)A_i^N(q_m,\boldsymbol{\kappa})\vector{e}_{\zeta} 
\end{equation}
for \(i \geq 0\),
\end{subequations}
with \(\boldsymbol{e}_{\zeta} = (0,0,1)^{\rm T}\) and the new introduced operator \(A^N_i\) defined by
\begin{equation}
	\label{eq:deviation_definition_pressure}
	A_i^N\left(q_m,\boldsymbol{\kappa}\right):= \frac{1}{h_0}(-1)^{i}q_m + \frac{1}{h_0}\sum_{j=1}^{N}\left(1 + \int_{0}^{1}\phi_i'\phi_j\,\mathrm{d}\zeta\right)\kappa_j.
\end{equation}
Remark, that \(A_i^N\) only enters the third equation that describes the evolution of the vertical velocity component \(w\).
The derivation of the equations~\eqref{eq:moments-evol} is mostly straightforward, since one can use the orthogonal property of the Legendre polynomials to determine the integrals. The only difficulty appears for the \(\zeta\)-derivative, what is the point where the operator \(A_i^N\) comes in. A full calculation is given below that also illustrates the weak enforcement of the boundary conditions~\eqref{eq:moment_boundary_surface}.
\begin{align*}
	\frac{1}{h_0}\int_{0}^{1}\partial_{\zeta}q\phi_i\,\mathrm{d}\zeta &= \frac{1}{h_0}\int_{0}^{1}\partial_{\zeta}\left(q_m+\sum_{j=1}^{N}\kappa_j\phi_j\right)\phi_i\,\mathrm{d}\zeta
	=\frac{1}{h_0}\sum_{j=1}^{N}\kappa_j\left(\left[\phi_j\phi_i\right]_0^1 -\int_{0}^{1}\phi_i'\phi_j\,\mathrm{d}\zeta\right)\\
	&=\frac{1}{h_0}\left(-q_m(-1)^i -\sum_{j=1}^{N}\kappa_j\left(1+\int_{0}^{1}\phi_i'\phi_j\,\mathrm{d}\zeta\right)\right).
\end{align*}
In a similar way we now consider the divergence constraint and get 
\begin{subequations}\label{eq:moments-div}
\begin{align}
\label{eq:moments_div_zero}
		&\hat{\nabla}\cdot \vector{u}_m =-B_0^N\left(w_m,\boldsymbol{\gamma}\right)  , \qquad\qquad i=0,\\
        \label{eq:moments_div_higher}
		&\hat{\nabla}\cdot\vector{\Lambda}_i =- (2i+1)B_i^N\left(w_m,\boldsymbol{\gamma}\right) ,\quad i\geq 1.
\end{align}
\end{subequations}
Again, we have calculated the \(\zeta\)-derivative separately and with that we have defined the operator~\(B_i^N(w_m,\boldsymbol{\gamma})\) via
\begin{equation}
	\label{eq:deviation_definition_vertvelocity}
	B_i^N\left(w_m,\boldsymbol{\gamma}\right):= \frac{1}{h_0}\left(w_m - \vector{u}_0\cdot\hat{\nabla}h_b +\sum_{j=1}^{N}\left((-1)^{i+j}-\int_{0}^{1}\phi_i'\phi_j\,\mathrm{d}\zeta\right)\gamma_j\right).
\end{equation}
Similar as for the pressure deviation before we get the result by first considering the definition of \(w\) and then performing integration by parts
\begin{align*}
	\frac{1}{h_0}\int_{0}^{1}\partial_{\zeta}w\phi_i\,\mathrm{d}\zeta
	=\frac{1}{h_0}\left(w_m-\vector{u}_0\cdot\hat{\nabla}h_b+\sum_{j=1}^{N}\gamma_j\left((-1)^{i+j}-\int_{0}^{1}\phi_j\phi_i'\,\mathrm{d}\zeta\right)\right),
\end{align*}
where we used the boundary condition~\eqref{eq:moment_boundary_bottom} for the vertical velocity at the bottom.
Notice that \(A_i^N\) and \(B_i^N\) do not depend on \(\zeta\) for all \(i=0,\dots,N\) such that the resulting equations are truly dimensionally reduced. 
The general Linear Dispersive Shallow Moment System of level \(N\) (LDSM-N) in two horizontal dimensions is given by the linear mass balance
\eqref{eq:lin_ref_sys_height}, the evolution equations~\eqref{eq:moments-evol} for the velocity means and moments  and the linear constraints~\eqref{eq:moments-div}.

\subsubsection{Pressure Equation for the Generic Linear Model (LDSM-N)}\label{subsubsec:pressure_eq_LDSM}

Like for the linear reference system~\eqref{eq:lin-ref-sys} it is possible to replace the divergence constraints~\eqref{eq:moments-div} with Poisson-like problems for the mean pressure deviation \(q_m\) and the higher order coefficients \(\kappa_j\). We proceed as in section~\ref{subsec:pressure_eq_lin_ref_sys} and take the divergence of evolution equations~\eqref{eq:moments-evol}; using the divergence constraints then yields the equations of the Poisson-like problem. We present the calculations for the averaged equation~\eqref{eq:moments_evol_zero}. Since the derivatives commute, we get after applying the divergence
\begin{equation}
\partial_t\left(\hat{\nabla}\cdot\vector{u}_m\right)+\left(\vector{u}_0\cdot\hat{\nabla}\right)\left(\hat{\nabla}\cdot\vector{u}_m\right) +g\hat{\Delta}h+\hat{\Delta}q_m=-g\hat{\Delta}h_b.
\end{equation}
In the first two terms we identify the divergence and substitute the right-hand side \(-B_0^N\) of the divergence constraints. Scaling by \(h_0\) allows us to reformulate the term. Here we give a calculation for the general case of \(B_i^N,\, i=0,\dots,N\).
\begin{align*}
	&h_0\left(\partial_tB_i^N\left(w_m,\boldsymbol{\gamma}\right) +\left(\vector{u}_0\cdot\hat{\nabla}\right)B_i^N\left(w_m,\boldsymbol{\gamma}\right)\right)\\
	&= \partial_tw_m +\sum_{j=1}^{N}\left((-1)^{i+j}-\int_0^1\phi_i'\phi_j\,\mathrm{d}\zeta\right)\partial_t\gamma_j \\
	&+\left(\vector{u}_0\cdot\hat{\nabla}\right)w_m - \left(\vector{u}_0\cdot\hat{\nabla}\right)\left(\vector{u}_0\cdot\hat{\nabla}h_b\right)+\sum_{j=1}^{N}\left((-1)^{i+j}-\int_0^1\phi_i'\phi_j\,\mathrm{d}\zeta\right)\left(\vector{u}_0\cdot\hat{\nabla}\right)\gamma_j\\
	&=A_0^N\left(q_m,\boldsymbol{\kappa}\right)-\left(\vector{u}_0\cdot\hat{\nabla}\right)\left(\vector{u}_0\cdot\hat{\nabla}h_b\right)+\sum_{j=1}^{N}\left((-1)^{i+j}-\int_0^1\phi_i'\phi_j\,\mathrm{d}\zeta\right)(2j+1)A_j^N\left(q_m,\boldsymbol{\kappa}\right)\\
	&=h_0\mathcal{B}_i^N\left(\left\{A_j^N\left(q_m,\boldsymbol{\kappa}\right)\right\}_j\right),
\end{align*}
where we have defined the operators \(\mathcal{B}_i^N\). The definition is very similar to the one of \(B_i^N\) 
\begin{equation}
    \label{eq:def_op_poisson}
    \mathcal{B}_i^N\left(\left\{A_j^N\right\}_j\right) = \frac{1}{h_0}\left( A_0^N-\left(\vector{u}_0\cdot\hat{\nabla}\right)\left(\vector{u}_0\cdot\hat{\nabla}h_b\right)+\sum_{j=1}^{N}\left((-1)^{i+j}-\int_0^1\phi_i'\phi_j\,\mathrm{d}\zeta\right)(2j+1)A_j^N\right).
\end{equation}
All in all this gives us an alternative formulation for our divergence constraint in form of a Poisson-like equation,
\begin{equation}
\label{eq:Poisson_lin_average}
    \hat{\Delta}(g(h+h_b)+q_m) = \mathcal{B}_0^N\left(\left\{A_j^N\left(q_m,\boldsymbol{\kappa}\right)\right\}_j\right).
\end{equation}
Notice that because of the terms involving $q_m$ and the $\kappa_k$ on the right hand side one cannot speak of a Poisson equation here. 
The calculation for the higher order equations \(i\geq 1\) is the same, replacing \(A_0^N\) by \((2i+1)A_i^N\) on the right-hand side does not change anything since the operator has no \(\zeta\)-dependency. We can again use the identity above after we have taken the divergence.\\
This gives the final form of the rewritten linear dispersive shallow moment system
\begin{subequations}\label{eq:lin-moments-poisson}
\begin{align}
\label{eq:lin_moments_poisson_height}
	&\partial_th + h_0\hat{\nabla}\cdot\vector{u}_m + \vector{u}_0\cdot\hat{\nabla}h =0,\\
    \label{eq:lin_moments_poisson_evol_average}
		&\partial_t\vector{u}_m + \left(\vector{u}_0\cdot\hat{\nabla}\right)\vector{u}_m +g\hat{\nabla}h +\hat{\nabla}q_m = -g\hat{\nabla}h_b + A_0^N\left(q_m,\boldsymbol{\kappa}\right)\vector{e}_{\zeta},\\
        \label{eq:lin_moments_poisson_evol_higher}
		&\partial_t\vector{\Lambda}_i +\left(\vector{u}_0\cdot\hat{\nabla}\right)\vector{\Lambda}_i + \hat{\nabla}\kappa_i = (2i+1)A_i^N\left(q_m,\boldsymbol{\kappa}\right)\vector{e}_{\zeta},\\
        \label{eq:lin_moments_poisson_average}
        &\hat{\Delta}(g(h+h_b)+q_m) = \mathcal{B}_0^N\left(\left\{A_{j}^N\left(q_m,\boldsymbol{\kappa}\right)\right\}_j\right),\\
        \label{eq:lin_moments_poisson_higher}
		&\hat{\Delta}\kappa_i = (2i+1)\mathcal{B}_i^N\left(\left\{A_j^N\left(q_m,\boldsymbol{\kappa}\right)\right\}_j\right),
\end{align}
for \(i=1,\dots,N\) and
with the two operators \(A_i^N\) and \(\mathcal{B}_i^N\) defined above in~\eqref{eq:deviation_definition_pressure} and~\eqref{eq:def_op_poisson}.
\end{subequations}
Remark that the operators \(B_i^N\) and \(\mathcal{B}_i^N\) are very similar and except for factors only differ on the number of derivatives of the bottom topography \(h_b\).

\subsection{Dispersive Shallow Moment Equations (DSM)}\label{subsec:DSM}

For modeling a fluid flow that cannot be interpreted as a small perturbation around a ground state with background velocity $u_0$ and fluid height $h_0$ the nonlinear equations come into play. Starting from the reference system~\eqref{eq:ref-sys} the derivation follows the known pattern: For a fixed number $N$ the velocity and pressure are replaced with a polynomial ansatz of length $N$ and the reduced system is derived with Galerkin projections.
Since the system was already derived in \cite{ScholzKowalskiTorrilhon_dispersionshallowmoment}, we restrict ourself to quickly go through the equations here with some comments on the new notation. A fully detailed derivation of the system can be found in the appendix~\ref{appx:subsec:DSM-N}.

 As in the linear case the height balance is already depth projected and  depends only on the averaged values of the velocity:
\begin{subequations}\label{eq:general_DSM}
\begin{align}
\label{eq:general_DSM_height}
        \partial_th + \partial_x(hu_m) + \partial_y(hv_m) = 0.
        \end{align}
        A fundamental difference between the linear and the nonlinear systems is that the nonlinear evolution equations are stated in terms of conservative variables. The evolution of the horizontal momentum \(h\boldsymbol{u}_m\)
        is given by
        \begin{align}
        \nonumber
        &\partial_t\left(h\vector{u}_m\right)+\partial_x\left(h\left(u_m\vector{u}_m+\sum_{j=1}^{N}\frac{\alpha_j\vector{\Lambda}_j}{2j+1}\right)\right)+\partial_y\left(h\left(v_m\vector{u}_m + \sum_{j=1}^{N}\frac{\beta_j\vector{\Lambda}_j}{2j+1}\right)\right)\\
        \label{eq:general_DSM_average_velo}
		&+\frac{g}{2}\tilde{\nabla}_{x}(h^2)+hg\tilde{\nabla}_{x}h_b +\tilde{\nabla}_{x}(hq_m) + \tilde{\nabla}_{x}h_b\left(q_m + \sum_{j=1}^{N}\kappa_j\right) - \left(q_m + \sum_{j=1}^{N}\kappa_j\right)\vector{e}_{\zeta} =0.
	\end{align}
    Thanks to the newly introduced operators the vertical velocity $w_m$ now seamlessly integrates with the other velocities regardless of its alignment with the directions of gravity and hydrostatic pressure.  Similarly we have for the evolution of the higher order velocity moments  \(\boldsymbol{\Lambda}_i\) 
	 \(i\geq 1\)
	\begin{align}
        \nonumber
		&\frac{\partial_t\left(h\vector{\Lambda}_i\right)}{2i+1}+\partial_x\left(h\left(\frac{u_m\vector{\Lambda}_i}{2i+1}+\frac{\alpha_i\vector{u}_m}{2i+1}+\sum_{j,k=1}^{N}\alpha_j\vector{\Lambda}_k\frac{A_{ijk}}{2i+1}\right)\right) \\&+ 
        \nonumber
        \partial_y\left(h\left(\frac{v_m\vector{\Lambda}_i}{2i+1}+\frac{\beta_i\vector{u}_m}{2i+1}+\sum_{j,k=1}^{N}\beta_j\vector{\Lambda}_k\frac{A_{ijk}}{2i+1}\right)\right)\\
        \nonumber
		&-\frac{\vector{u}_m\tilde{\nabla}_x\cdot(h\boldsymbol{\Lambda}_i)}{2i+1}+\sum_{j,k=1}^{N}\vector{\Lambda}_k\tilde{\nabla}_x\cdot(h\boldsymbol{\Lambda}_j)\frac{B_{ijk}}{2i+1}+\frac{h\tilde{\nabla}_{x}\kappa_i}{2i+1}\\
        \label{eq:general_DSM_higher_moments_velo}
        &-\tilde{\nabla}_{x}h\sum_{j=1}^{N}\frac{\kappa_jH_{ji}}{2j+1}
		-\tilde{\nabla}_{x}h_b\sum_{j=1}^{N}\frac{\kappa_jG_{ji}}{2j+1}+\sum_{j=1}^{N}\kappa_j\frac{G_{ji}}{2j+1}\vector{e}_{\zeta}=0,
	\end{align}
    as one single vectorial equation for all directions of velocity.
    
    The depth-averaged constraint becomes
        \begin{align}
        \label{eq:general_DSM_average_constraint}
            h\tilde{\nabla}_{x}\cdot\vector{u}_m + w_m -\vector{u}_m\cdot\tilde{\nabla}_{x}h_b +\sum_{j=1}^{N}\left(\gamma_j(-1)^j 
		-\tilde{\nabla}_{x}h\cdot\vector{\Lambda}_j (-1)^j - \tilde{\nabla}_{x}h_b\cdot\vector{\Lambda}_j(-1)^j \right)=0.
        \end{align}
        Unlike the above evolution equations, this is a scalar equation as there is only one scalar divergence constraint from which the constraint is deduced. The Legendre-projections of the constraint are finally
\begin{align}
        \nonumber
        &\frac{h\tilde{\nabla}_{x}\cdot\vector{\Lambda}_i}{2i+1}+w_m-\vector{u}_m\cdot\tilde{\nabla}_{x}h_b-\sum_{j=1}^{N}\left(\vector{\Lambda}_j\cdot\tilde{\nabla}_{x}h_b+\gamma_j\left((-1)^{i+j}-\frac{G_{ij}}{2i+1}\right)\right)\\
        \label{eq:general_DSM_higher_moments_constraint}
		&-\sum_{j=1}^{N}\left(\tilde{\nabla}_{x} h \cdot\vector{\Lambda}_j \frac{H_{ji}}{2j+1}- \tilde{\nabla}_{x}h_b\cdot\vector{\Lambda}_j\frac{G_{ji}}{2j+1}\right)=0.
\end{align}
    A definition of the abbreviations \(A,B,G\) and \(H\) can be found in the appendix~\ref{appx:subsubsec:abbr}.
\end{subequations}

\subsubsection{Remark on the Reformulation of the Pressure Constraint}\label{sec:reformulation_nonlinear_constraint}

Transforming the divergence constraint in the nonlinear reference system and the nonlinear moment systems into a Poisson-like problem in an elegant way remains difficult. However, the transformation can be done by clever manipulation of the equations. The necessary steps are exemplified here in a schematic way in one spatial dimension. With the help of a computer algebra system the manipulation is done semi-automatically for the nonlinear systems since the equations that arise are difficult to handle due to their length and high number of terms. This can be observed even for the simplest system of level zero in subsection~\ref{subsec:example_DSM0}. The complexity of the equations even in the simplified form, increases rapidly. An explicit calculation of the pressure equation for the zeroth level system is given in the appendix~\ref{appx:subsec:pressure_constraint_dsm0}.\\
The general DSM system in one dimension is obtained by the equations~\eqref{eq:general_DSM} from the section before restricted to one horizontal dimension by neglecting all \(y\) derivatives and setting \(v_m\) as well as all higher moments \(\beta_i\) to zero. To reformulate the divergence constraints~\eqref{eq:general_DSM_average_constraint} and~\eqref{eq:general_DSM_higher_moments_constraint} we perform the following steps:
\begin{enumerate}
    \item The evolution equations are reformulated in pure variables and the height balance \eqref{eq:general_DSM_height} is applied. This yields
    \begin{align}     \label{eq:general_solved_v0_vec}
         \partial_t \begin{pmatrix}
             u_m \\ \boldsymbol{\alpha}
         \end{pmatrix} &=\frac{1}{h}\left(\partial_x(h u_m)\begin{pmatrix}
             u_m \\ \boldsymbol{\alpha}
         \end{pmatrix} + \begin{pmatrix}
            U^N \\ \boldsymbol{K}^N
        \end{pmatrix}\right),\\
        \label{eq:general_solved_v1_vec}
         \partial_t \begin{pmatrix}
             w_m \\ \boldsymbol{\gamma}
         \end{pmatrix} &=\frac{1}{h}\left(\partial_x(h u_m)\begin{pmatrix}
             w_m \\ \boldsymbol{\gamma}
         \end{pmatrix} + \begin{pmatrix}
            W^N \\ \boldsymbol{\Gamma}^N
        \end{pmatrix}\right).
    \end{align}
    with \(\boldsymbol{\alpha} = (\alpha_1,\dots,\alpha_N)^{\rm T}, \, \boldsymbol{\gamma}= (\gamma_1,\dots,\gamma_N)^{\rm T}\) as well as \(\boldsymbol{K}^N = \left(3K_1^N, 5K_2^N,\dots, (2N+1)K_N^N\right)^{\rm T}\) and \(\boldsymbol{\Gamma}^N = \left(3\Gamma_1^N,\dots, (2N+1)\Gamma_N^N\right)^{\rm T}\).
    Here, the introduced operators \(U^N, W^N, \boldsymbol{K}^N,\boldsymbol{\Gamma}^N\) combine their arguments by summation, multiplication and constant factors only. Remark, that the operators have different dependencies corresponding to the underlying equation given by~\eqref{eq:general_DSM_average_velo} and~\eqref{eq:general_DSM_higher_moments_velo}. In fact, these dependencies are:
\begin{subequations}
\begin{align}
    U^N\left(h,u_m, q_m,\boldsymbol{\alpha}, \boldsymbol{\kappa},\partial_xh,\partial_xu_m,\partial_xq_m,\partial_x\boldsymbol{\alpha},\partial_xh_b\right),\\
    W^N\left(h,u_m, w_m, q_m,\boldsymbol{\alpha},\boldsymbol{\gamma}, \boldsymbol{\kappa},\partial_xh,\partial_xu_m,\partial_xw_m,\partial_x\boldsymbol{\alpha},\partial_x\boldsymbol{\gamma}\right),\\
    K^N_i\left(h,u_m,\boldsymbol{\alpha},\boldsymbol{\kappa},\partial_xh,\partial_xu_m,\partial_x\boldsymbol{\alpha},\partial_xh_b\right),\\
    \Gamma^N_i\left(h,u_m, w_m, q_m,\boldsymbol{\alpha},\boldsymbol{\gamma},\boldsymbol{\kappa},\partial_xh,\partial_xu_m,\partial_xw_m,\partial_x\boldsymbol{\alpha},\partial_x\boldsymbol{\gamma}\right),
\end{align}
where we have in addition \(\boldsymbol{\kappa} = (\kappa_1,\dots,\kappa_N)^{\rm T}\).
\end{subequations}
\item The constraints \eqref{eq:general_DSM_average_constraint} and~\eqref{eq:general_DSM_higher_moments_constraint} combined take the general vectorial form
\begin{equation}
    h \partial_x \boldsymbol{V}_0+A\boldsymbol{V}_0 \partial_x h+ B \boldsymbol{V}_1+C\boldsymbol{V}_0 \partial_x h_b=0,
\end{equation}
where $\boldsymbol{V}_0=(u_m,\boldsymbol{\alpha}^{\rm T})^{\rm T}, \boldsymbol{V}_1=(w_m, \boldsymbol{\gamma}^{\rm T})^{\rm T}$ are variable vectors and $A,B,C$ are real coefficient matrices.
Consequently, the temporal derivative of the constraints is
\begin{equation}
\label{eq:temporal_constraint}
    \partial_t h \partial_x \boldsymbol{V}_0+h \partial_x \partial_t \boldsymbol{V}_0+A \partial_t \boldsymbol{V}_0 \partial_x h+A \boldsymbol{V}_0 \partial_x \partial_t h+B \partial_t \boldsymbol{V}_1 + C \partial_t \boldsymbol{V}_0 \partial_x h_b=0.
\end{equation}
    \item Similarly, the spatial derivative of the constraints reads 
    \begin{equation}
    \label{eq:spatial_constraint}
        \partial_xh\partial_x\boldsymbol{V}_0+h\partial_{xx}\boldsymbol{V}_0+A \partial_x\boldsymbol{V}_0 \partial_xh+A\boldsymbol{V}_0\partial_{xx}h+B\partial_x\boldsymbol{V}_1+C\partial_x\boldsymbol{V}_0\partial_xh_b+C\partial_{xx}h_b.
    \end{equation}
    \item In the last step equation \eqref{eq:temporal_constraint} multiplied by the height $h$ and \eqref{eq:spatial_constraint} multiplied by the momentum $h u_m$ are added up. The expressions $\partial_t \boldsymbol{V}_0$ and $\partial_t \boldsymbol{V}_1$ and $\partial_t h$ can be replaced using \eqref{eq:general_solved_v0_vec} and \eqref{eq:general_solved_v1_vec} and the height balance \eqref{eq:general_DSM_height}. The resulting equation is the Poisson-like problem
    which can be brought into a more readable form using a computer algebra system.
\end{enumerate}

\section{Examples of Moment Systems}\label{sec:examples_moment_systems}

After showing how the general systems are derived, this section presents linear and nonlinear moment systems for the levels \(N=0\) until \(N=3\). Like \cite{ScholzKowalskiTorrilhon_dispersionshallowmoment} we present the systems in dimensionless variables. To achieve this, we use the scaling
\begin{equation}
\label{eq:scaled_variables}
\hat h = \frac{h}{h_0},\quad \hat u = \frac{u}{u_0},\quad \hat w = \frac{w}{w_0},\quad \hat q=\frac{q}{q_0},\quad \hat h_b = \frac{h_b}{l_0},\quad \hat x=\frac{x}{l_0},\quad \hat t= \frac{u_0 t}{l_0},
\end{equation} 
with the characteristic length $l_0$, height $h_0$ and velocity $u_0$
as well as the derived quantities $w_0=u_0h_0/l_0
$ and $q_0=g h_0$.
With this quantities we define the Froude number $\Fr = u_0/\sqrt{g h_0}$ and the shallowness parameter $S=h_0/l_0$ which
are the dimensionless flow characteristics. To increase readability we omit the hat on the variables and agree on the use the dimensionless variables for the rest of this section. For simplicity we restrict ourself to one horizontal direction in the following particular moment systems.

\subsection{LDSM0}\label{subsec:example_LDSM0}

The simplest element of the LDSM cascade is the level zero, or LDSM0 system. It is given by the equations
\begin{subequations}\label{eq:LDSM0}
\begin{align}
\label{eq:LDSM0_evol}
&\partial_t\begin{pmatrix}
    h \\ \Fr^2 u_m \\ \He^2\Fr^2 w_m
\end{pmatrix}+ \partial_x\begin{pmatrix}
    h + u_m \\ \Fr^2 u_m + h + q_m \\ \He^2 \Fr^2 w_m
\end{pmatrix} = \begin{pmatrix}
    0 \\ 0 \\ q_m
\end{pmatrix} - \frac{1}{\He}\partial_xh_b\begin{pmatrix}
    0 \\ 1 \\ 0
\end{pmatrix},\\
\label{eq:LDSM_constraint}
&\dx u_m = \frac{1}{\He} \dx h_b - w_m,
\end{align}
where the constraint~\eqref{eq:LDSM_constraint} can equivalently be replaced by the Poisson-like problem
\begin{equation}
    \label{eq:LDSM0_poisson}
    \partial_{xx}q_m = \frac{1}{\He^2}q_m -\partial_{xx}\left(h + \frac{1}{\He}h_b\right) - \frac{\Fr^2}{\He}\partial_{xx}h_b.
\end{equation}
\end{subequations}
following the procedure of \ref{subsubsec:pressure_eq_LDSM}.
The constraint in the Poisson-like form does depend on the height $h$, but not the velocities. In the case of $\partial_x h=0$, like in a lake of rest state, the non-hydrostatic pressure is entirely determined through the shape of the bottom topography $h_b$.
We keep in mind that the total (unmapped) pressure is $p=p_{hyd}+q=g(h+h_b-z)+q$ which gives the relation $hp_m=hq_m+gh(h+h_b)-\frac{g h^2}{2}-g h h_b$ when integrated $\int_{h_b}^{h_s}\,\mathrm{d} z$, leading to the relation $p_m=\frac{g h}{2}+q_m$. The Poisson-like problem could therefore also be formulated in terms of $p_m$, possibly bringing numerical advantages. The investigation remains further work.

\subsection{LDSM1}\label{subsec:example_LDSM1}

While the LDSM0 system is primarily useful for gaining an impression of the structure of the linear moment systems, the LDSM1 system is already relevant in practical terms. It is a simple linear system that models the horizontal velocity as a first-order polynomial and approximates the full system in the dispersion relation up to an order of accuracy of $K^3$ as shown in \cite{ScholzKowalskiTorrilhon_dispersionshallowmoment}.
\begin{subequations}\label{eq:LDSM1}
    \begin{align}
        \label{eq:LDSM1_evol}
        \partial_t\begin{pmatrix}
            h \\ \Fr^2 u_m \\ \Fr^2 \alpha_1 \\ \He^2\Fr^2 w_m \\ \He^2\Fr^2 \gamma_1
        \end{pmatrix} &+ \partial_x\begin{pmatrix}
            h + u_m \\ \Fr^2 u_m + h + q_m \\ \Fr^2 \alpha_1 + \kappa_1 \\ \He^2\Fr^2 w_m \\ \He^2 \Fr^2 \gamma_1
        \end{pmatrix} = \begin{pmatrix}
            0 \\ 0 \\ 0 \\ q_m + \kappa_1 \\ -3q_m + 3\kappa_1
        \end{pmatrix} - \frac{1}{\He}\partial_xh_b\begin{pmatrix}
            0 \\ 1 \\ 0 \\ 0 \\ 0 
        \end{pmatrix}\\
        \label{eq:LDSM1_constraint_zeroth}
        &\partial_xu_m = \frac{1}{\He}\partial_xh_b + \gamma_1 - w_m,\\
        \label{eq:LDSM1_constraint_first}
        &\partial_x\alpha_1 = \frac{3}{\He}\partial_xh_b - 3w_m - 3 \gamma_1.
    \end{align}
    Again we can replace the constraints by a Poisson-like problems with the average pressure deviation \(q_m\) and the first order coefficient \(\kappa_1\) as unknown variables. This yields
    \begin{align}
        \label{eq:LDSM1_poisson_zeroth}
        & \partial_{xx}q_m = \frac{4}{\He^2}q_m - \frac{2}{\He^2}\kappa_1 -\partial_{xx}\left(h + \frac{1}{\He}h_b\right)- \frac{\Fr^2}{\He}\partial_{xx}h_b,\\
        \label{eq:LDSM1_poisson_first}
        &\partial_{xx}\kappa_1 = -\frac{6}{\He^2}q_m + \frac{12}{\He^2}\kappa_1 - \frac{3\Fr^2}{\He}\partial_{xx}h_b.
    \end{align}
\end{subequations}
The first constraint \eqref{eq:LDSM1_poisson_zeroth} is similar in structure to \eqref{eq:LDSM0_poisson} but is coupled to the second constraint \eqref{eq:LDSM1_poisson_first} through the pressure moment $\kappa_1$.

\subsection{LDSM2}\label{subsec:example_LDSM2}

The next element of the LDSM cascade and the last one we will state here is the LDSM2 system which gives an impression of how the systems continue for higher levels.
\begin{subequations}\label{eq:LDSM2}
    \begin{align}
        \label{eq:LDSM2_evol}
           \partial_t\begin{pmatrix}
               h \\ \Fr^2u_m \\ \Fr^2 \alpha_1 \\ \Fr^2\alpha_2 \\ \He^2\Fr^2 w_m \\ \He^2\Fr^2 \gamma_1 \\ \He^2\Fr^2\gamma_2
           \end{pmatrix} &+ \partial_x\begin{pmatrix}
               h + u_m \\ \Fr^2u_m + h + q_m \\ \Fr^2 \alpha_1 + \kappa_1 \\ \Fr^2 \alpha_2 + \kappa_2 \\ \He^2\Fr^2 w_m \\ \He^2\Fr^2 \gamma_1 \\ \He^2 \Fr^2 \gamma_2
           \end{pmatrix}  = \begin{pmatrix}
               0 \\ 0 \\ 0 \\ 0 \\ q_m + \kappa_1 + \kappa_2 \\ -3q_m + 3\kappa_1 + 3\kappa_2 \\ 5q_m - 5\kappa_1 + 5\kappa_2 
           \end{pmatrix} - \frac{1}{\He}\partial_xh_b\begin{pmatrix}
               0 \\ 1 \\ 0 \\ 0 \\ 0 \\ 0 \\ 0
           \end{pmatrix}, \\
        \label{eq:LDSM2_constraint_zeroth}
       &\partial_x u_m = \frac{1}{\He}\partial_xh_b - w_m + \gamma_1 - \gamma_2,\\
        \label{eq:LDSM2_constraint_first}
        &\partial_x\alpha_1 = \frac{3}{\He}\partial_xh_b - 3w_m -3\gamma_1 + 3\gamma_2,\\
        \label{eq:LDSM2_constraint_second}
        &\partial_x\alpha_2 = \frac{5}{\He}\partial_xh_b - 5w_m - 5\gamma_1 - 5\gamma_2,
    \end{align}
    As before we can replace the divergence constraints with Poisson-like equations for the pressure variables.
    \begin{align}
        \label{eq:LDSM2_poisson_zeroth}
        \partial_{xx}q_m &= \frac{9}{\He^2}q_m - \frac{7}{\He^2}\kappa_1 + \frac{3}{\He^2}\kappa_2 -\partial_{xx}\left(h + \frac{1}{\He}h_b\right) - \frac{\Fr^2}{\He}\partial_{xx}h_b,\\
        \label{eq:LDSM2_poisson_first}
        \partial_{xx}\kappa_1 &= -\frac{21}{\He^2}q_m + \frac{27}{\He^2}\kappa_1 - \frac{3}{\He^2}\kappa_2 - \frac{3\Fr^2}{\He}\partial_{xx}h_b,\\
        \label{eq:LDSM2_poisson_second}
        \partial_{xx}\kappa_2 &= \frac{15}{\He^2}q_m - \frac{5}{\He^2}\kappa_1 + \frac{45}{\He^2}\kappa_2 - \frac{5\Fr^2}{\He}\partial_{xx}h_b.
    \end{align}
\end{subequations}
The Poisson-like system has the same structure as that of LDSM0 and LDSM1.

As discussed in \cite{ScholzKowalskiTorrilhon_dispersionshallowmoment}, the (L)DSM systems have a hyperbolic core from the existing evolution equations for height and the velocities. The hyperbolicity properties of this core was discussed in \cite{KowalskiTorrilhon_shallowmomentapprox} where it was shown that the level two system is only locally hyperbolic. The linear systems, however, all take the form
\begin{align}
    \partial_t \boldsymbol{V}+A_{sys} \partial_x \boldsymbol{V}=-\boldsymbol{P}; \qquad
    A_{sys}=\begin{pmatrix}
        1 & 1 & \mathbf0\\
        \frac{1}{Fr^2} & 1 & \mathbf0 \\
        \mathbf0& \mathbf0 &  I
    \end{pmatrix}
\end{align}
where $A_{sys}$ has $N$ independent eigenvectors making any linear LDSM system globally hyperbolic. Linearization can serve the purpose of hyperbolic regularization, although there are more advanced and less restrictive options \cite{KoellermeierRominger2020}.
Now the nonlinear systems will be discussed briefly.

\subsection{Level zero system or extended shallow water equations}\label{subsec:example_DSM0}

As discussed in \cite{ScholzKowalskiTorrilhon_dispersionshallowmoment}, the DSM0 system
    \begin{subequations}\label{example:dsm0}
    \begin{equation}
    \label{eq:dsm0}
        \dt
        \begin{pmatrix}
            h \\
            \Fr^2 h u_m \\
            \He^2 \Fr^2 h w_m
        \end{pmatrix}
        +
        \dx
        \begin{pmatrix}
            h u_m \\
            \frac{h^2}{2}+ \Fr^2 h u_m^2 +h q_m\\
            \He^2 \Fr^2 h u_m w_m
        \end{pmatrix}
        =
        \begin{pmatrix}
            0\\
            0 \\
            q_m\\
        \end{pmatrix} - \frac{1}{\He} \dx h_b
        \begin{pmatrix}
            0\\
             h+q_m \\
            0 \\
        \end{pmatrix}\hspace{45pt}\\[4ex]
\end{equation}
has one single constraint
\begin{equation}
\label{eq:dsm0_constr}
        \hfill h \dx u_m +w_m - \frac{1}{\He} u_m \dx h_b = 0
    \end{equation}
    obtained from the divergence free condition.
    The corresponding Poisson-like equation following the procedure of section \ref{sec:reformulation_nonlinear_constraint} is
    \begin{align}
    \begin{split}
    \label{eq:dsm0_poisson}
    h^2 \partial_{xx}q_m=&-\left(hq_m+h^2\right)\partial_{xx}h- \left(\frac{1}{\He} hq_m+\frac{\Fr^2}{\He} hu_m^2+\frac{1}{\He} h^2\right)\partial_{xx} h_b\\&+
        \frac{1}{\He} 2 q_m \dx h_b \partial_x h
        +\frac{1}{\He} h \dx h_b \partial_x h 
        -h \partial_x h  \partial_x q_m
    +q_m (\partial_x h)^2 \\
    &+\frac{1}{\He^2} h (\partial_x h_b)^2 
    -2 \Fr^2 h^2 (\partial_x u_m)^2
    +\frac{1}{\He^2} q_m(\dx h_b)^2+\frac{1}{\He^2} q_m
    \end{split}
    \end{align}
    \end{subequations}
Structurally, this constraint differs from its linear counterpart \eqref{eq:LDSM0_poisson} in the direct dependency on the horizontal velocity $u_m$ and the non-constant factors of the unknowns $q_m$, $\partial_x q_m$ and $\partial_{xx} q_m$. This motivates describing the constraint as quasi-linear. Numerical splitting, however, will allow for interpretation as a linear PDE.
For brevity, only the equations of the Poisson-like problem for the systems DSM0 and DSM1 are explicitly stated.

\subsection{Level one system}\label{subsec:example_DSM1}

Starting from the equations
\begin{subequations}\label{example:dsm1}
    \begin{equation}
    \label{eq:dsm1}
        \dt
        \begin{pmatrix}
            h \\
            \Fr^2 h u_m \\
            \Fr^2 h \alpha_1 \\
            \He^2 \Fr^2 h w_m \\
            \He^2 \Fr^2 h \gamma_1 \\
        \end{pmatrix}
        +
        \dx
        \begin{pmatrix}
            h u_m \\
            \frac{h^2}{2}+\Fr^2(h u_m^2 +\frac{\alpha_1^2}{3}) +h q_m\\
            2 \Fr^2 h u_m\alpha_1 +h \kappa_1\\
            \He^2 \Fr^2 h( u_m w_m+ \frac{\alpha_1 \gamma_1}{3}) \\
            \He^2 \Fr^2 h (\alpha_1 w_m +  u_m \gamma_1) \\
        \end{pmatrix}
        =Q \partial_x
        \begin{pmatrix}
        h\\
        \Fr^2 h u_m \\
        \Fr^2 h \alpha_1 \\
        \He^2 \Fr^2 h w_m \\
        \He^2 \Fr^2h \gamma_1
        \end{pmatrix} -\boldsymbol{P}
        \end{equation}
        with
        \begin{equation}
        \boldsymbol{P}=
        \begin{pmatrix}
            0 \\
            0 \\
            0 \\
            -q_m - \kappa_1 \\
            3(q_m - \kappa_1) \\
        \end{pmatrix} +\frac{1}{\He}\partial_xh_b
        \begin{pmatrix}
            0 \\
            h+ q_m + \kappa_1\\
            0 \\
           0 \\
           0 \\
        \end{pmatrix}
    \text{ and }
    Q=
    \begin{pmatrix}
    0 &0 &0 &0 &0 \\
    0 &0 &0 &0 &0 \\
    2 \kappa_1 & 0& \Fr^2 u_m& 0& 0\\
    0 &0 &0 &0 &0 &\\
    0& 0& \Fr^2 \He^2 w_m& 0& 0
    \end{pmatrix}
    \end{equation}
    and the constraints
        \begin{align}
         &h \partial_x u_m + \alpha_1 \partial_x h +w_m -\gamma_1 - \frac{1}{\He} (u_m -\alpha_1) \partial_xh_b = 0 \text{ and } \\
         &h \partial_x \alpha_1 - \alpha_1 \partial_x h  +3 (w_m + \gamma_1) - \frac{3}{\He}(u_m + \alpha_1)\partial_x h_b = 0
        \end{align}
        a reformulation yields the following Poisson-like problem consisting of one equation for the non-hydrostatic pressure mean:
\begin{align}
h^2 &\partial_{xx}{q_m}=\\ \nonumber
&-\Fr^2\alpha_1h^2
   \partial_{xx}{u_m}-
\frac{2}{3}  \Fr^2h^2 \alpha_1
   \partial_{xx}\alpha_1 -\left(\frac{1}{3}  \Fr^2 \alpha_1^2 h+h^2+h q_m\right)
   \partial_{xx}{h}\\ \nonumber
   &+\frac{ \Fr^2}{ \He} \alpha_1h
   u_m\partial_{xx}h_b-\frac{ \Fr^2}{ \He}  h u_m^2\partial_{xx}h_b-\frac{1}{\He}\left(
   h\kappa_1- h
   q_m- h^2\right)\partial_{xx}h_b
    -\frac{1}{3}  \Fr^2 \gamma_1
   h \partial_x\alpha_1\\ \nonumber&-\frac{2}{3}  \Fr^2 h \alpha_1\partial_x\alpha_1
    \partial_x{h}+\frac{2  \Fr^2}{3  \He} 
    h \alpha_1
   \partial_x\alpha_1 \partial_{x}h_b  -\frac{2}{3}  \Fr^2 h^2
   (\partial_{x}\alpha_1)^2 -\frac{1}{3}  \Fr^2 h \alpha_1
   \partial_x{\gamma_1} \\ \nonumber
   &-\frac{1}{3}  \Fr^2 \alpha_1
  \gamma_1 \partial_x{h}\!+\frac{ \Fr^2}{3  \He} 
   \alpha_1^2 \partial_x{h} \partial_{x}h_b
   -3  \Fr^2\alpha_1h
   \partial_x h \partial_x{u_m}+\!\frac{1}{3}  \Fr^2 \alpha_1^2
   (\partial_{x}h)^2-\!\frac{ \Fr^2}{ \He} \alpha_1h
   \partial_x{u_m}\partial_{x}h_b\\ \nonumber
   &+ \Fr^2\alpha_1h
   \partial_x{w_m}-2
    \Fr^2 h^2 (\partial_{x}u_m)^2+\frac{1}{ \He}( 2\kappa_1  \partial_x{h}
  \partial_{x}h_b+2 q_m  \partial_x{h}
   \partial_{x}h_b+ h \partial_x{h}\partial_{x}h_b)\\ \nonumber&-h
   \partial_x{h} \partial_x{\kappa_1}
   +\kappa_1 (\partial_{x}h)^2
  -h \partial_x{h} \partial_x{q_m}+q_m (\partial_{x}h)^2
   +\frac{1}{ \He^2}h(\partial_{x}h_b)^2
   -\frac{2}{\He} h \partial_x{\kappa_1} \partial_{x}h_b \\ \nonumber
   &+\frac{1}{\He^2}\left(\kappa_1 (\partial_{x}h_b)^2
  +q_m (\partial_{x}h_b)^2 -2
  \kappa_1+4 q_m\right)
\end{align}
and another equation for the first non-hydrostatic pressure moment
\begin{align}
h^2 &\partial_{xx}{\kappa_1}=\\ \nonumber&
h  \kappa_1 \partial_{xx}{h}-\frac{3 \Fr^2}{\He} \alpha_1 h u_m \partial_{xx}{h_b}
   -\frac{3 \Fr^2}{\He}h   u_m^2 \partial_{xx}{h_b} 
\\ \nonumber&- \Fr^2 \gamma_1 h\partial_x{\alpha_1} +\frac{2 \Fr^2}{\He}
   h \alpha_1 \partial_x{\alpha_1} \partial_{x}{h_b}-3
   \Fr^2 h^2 \partial_x{\alpha_1} \partial_x{u_m}-\Fr^2
   h \alpha_1 \partial_x{\gamma_1} 
   -\Fr^2 \alpha_1
   \gamma_1 \partial_x{h}\\ \nonumber&+\frac{\Fr^2}{\He} 
   \alpha_1^2 \partial_x{h}\partial_{x}{h_b}+3 \Fr^2 \alpha_1 h
   \partial_x{h} \partial_x{u_m}+\!\frac{3 \Fr^2}{\He}
   \alpha_1 h \partial_x{u_m} \partial_{x}{h_b}
   -\!3 \Fr^2 \alpha_1
   h \partial_x{w_m}-\frac{3}{\He} \kappa_1  \partial_x{h}
   \partial_{x}{h_b}\\ \nonumber&+\frac{3}{\He}q_m \partial_x{h}  \partial_{x}{h_b}
   +\frac{3}{\He}
    h \partial_x{h}\partial_{x}{h_b}+2 h \partial_x{h}
   \partial_x{\kappa_1}-2 \kappa_1 (\partial_{x}h)^2 +\frac{3}{\He^2} h (\partial_{x}{h_b})^2 +\frac{3}{\He}
    h \partial_x{\kappa_1}\partial_{x}{h_b}\\ \nonumber
   &+\frac{3}{\He}
   h \partial_x{q_m} \partial_{x}{h_b}+\frac{3}{\He^2}
    \kappa_1 (\partial_{x}{h_b})^2+\frac{3}{\He^2}q_m (\partial_{x}{h_b})^2
   +\frac{12}{\He^2} \kappa_1-\frac{6}{\He^2} q_m.
\end{align}
\end{subequations}
Noticeable is the direct dependency of the reformulated constraint on the vertical velocity variables $w_m$ and $\gamma_1$ which marks another difference between the linear and the nonlinear case.
A linearization of these dimensionless constraints around $h=u_m=1$ and $\alpha_i=w_m=\gamma_i=q_m=\kappa_i=0$ leads to the constraints of the linear system \eqref{eq:LDSM1_poisson_zeroth} and \eqref{eq:LDSM1_constraint_first}.

\section{Numerical Experiments}\label{sec:Num_Experiments}

\subsection{Variable vector and spatial discretization}

For the linear LDSM systems, the variable vector $\boldsymbol{V}$ assembles the height $h$, the horizontal velocity mean $u_m$, the corresponding moments $\{\alpha_i\}_i$, the variables for the vertical velocity $w_m$ and $\{\gamma_i\}_i$ and finally the pressure variables $q_m$ and $\{\kappa_i\}_i$.
For the nonlinear systems, the vector contains the conservative variables $h$, $hu_m$, $\{h\alpha_i\}_i$, $hw_m$, $ \{h\gamma_i\}_i$ and the pressure variables $q_m$ and $\{\kappa_i\}_i$.
 An equidistant grid with $n_x$ cells is used for the 1D discretization. Consequently, the cell averages are denoted \begin{equation}
 \boldsymbol{V}_k=(h_k,u_{m,k},\alpha_{1,k},\dots,w_{m,k},\gamma_{1,k},\dots,q_{m,k},\kappa_{1,k},\dots),
 \end{equation} 
 for $k=1,\dots,n_x$ in pure variables. The notation for the conservative variables is similar.
 
 \subsection{The Splitting Scheme}\label{subsec:projection_method}

Let $S(\cdot)$ denote the solution operator associated to the full LDSM system \eqref{eq:lin-moments-poisson} or DSM system \eqref{eq:general_DSM}.
 Following the notation of \cite{Bonneton2011}, an approximation $\boldsymbol{V}^{n+1}$ at time $t_{n+1}=t_n+\delta_t$ can be found in terms of the approximation $\boldsymbol{V}^n$ at time $t_n$ by solving 
\begin{equation}
\label{eq:time-step}
\boldsymbol{V}^{n+1}=S(\delta_t)\boldsymbol{V}^n.
\end{equation}
All (L)DSM systems are composed of transport equations \begin{itemize}
        \item a mass balance \eqref{eq:lin_moments_poisson_height}, or, in the nonlinear case, \eqref{eq:general_DSM_height},
        \item balance equations \eqref{eq:lin_moments_poisson_evol_average}/\eqref{eq:general_DSM_average_velo} for the averages of horizontal and vertical velocity
        \item and balance equations \eqref{eq:lin_moments_poisson_evol_higher}/\eqref{eq:general_DSM_higher_moments_velo} for the higher moments of the velocities
    \end{itemize}  
  and constraints
    \begin{itemize}
        \item for the mean of the non-hydrostatic pressure \eqref{eq:lin_moments_poisson_average}/\eqref{eq:general_DSM_average_constraint} as well as
        \item for the respective moments \eqref{eq:lin_moments_poisson_higher}/\eqref{eq:general_DSM_higher_moments_constraint}.
    \end{itemize}
These coupled components are now separated, in other words, the operator $S(\cdot)$ is split at each time step $\delta_t$ in the following way:
\begin{equation}
    \label{eq:solution-split}
    S(\delta_t)=S_1(\delta_{t})S_2(\delta_{t}),
\end{equation}
where $S_2$ is the solution operator associated to the transport equations augmented with $\partial_t q_m= \partial_t \kappa_i=0$ and $S_1$ is associated to the pressure constraints. Precisely: 
\begin{itemize}
    \item For LDSM, $S_2$ is the solution operator associated to the altered LDSM system
    \begin{subequations}
    \begin{align}
	&\partial_th + h_0\hat{\nabla}\cdot\vector{u}_m + \vector{u}_0\cdot\hat{\nabla}h =0,\\
		&\partial_t\vector{u}_m + \left(\vector{u}_0\cdot\hat{\nabla}\right)\vector{u}_m +g\hat{\nabla}h +\hat{\nabla}q_m = -g\hat{\nabla}h_b + A_0^N\left(q_m,\boldsymbol{\kappa}\right)\vector{e}_{\zeta} ,\quad i=0\\
		&\partial_t\vector{\Lambda}_i +\left(\vector{u}_0\cdot\hat{\nabla}\right)\vector{\Lambda}_i + \hat{\nabla}\kappa_i = (2i+1)A_i^N\left(q_m,\boldsymbol{\kappa}\right)\vector{e}_{\zeta} ,\quad i\geq 1,\\
        &\partial_t q_m=0,\\
		&\partial_t \kappa_i = 0.
\end{align}
\end{subequations}
 \item For DSM, $S_2$ is the solution operator associated to the respective non-linear system \begin{equation}
     \{
     \eqref{eq:general_DSM_height},
     \eqref{eq:general_DSM_average_velo},
     \eqref{eq:general_DSM_higher_moments_velo},
     \partial_t q_m=0,
     \partial_t \kappa_i = 0\}.
 \end{equation}
\item For LDSM, $S_1$ is associated to the linear constraints
$\{\eqref{eq:lin_moments_poisson_average},\eqref{eq:lin_moments_poisson_higher}\}$.
\item For DSM, $S_1$ is associated to the constraints $\{\eqref{eq:general_DSM_average_constraint}, \eqref{eq:general_DSM_higher_moments_constraint}\}$. 
\end{itemize}
The splitting of the solution operator \eqref{eq:solution-split} means that two separate steps are performed at every timestep: First, values for the height and the velocity variables are calculated while the pressure is treated as input data. Then, the height and velocity variables are fixed and the pressure is updated.

\subsubsection{The hyperbolic step \texorpdfstring{$S_2$}{S2}}

This step computes $S_2(\delta_t)\boldsymbol{V}^n$. The equation to solve takes the form
\begin{align}
    \label{eq:general-transport}
    \partial_t \boldsymbol{V} + \partial_x \boldsymbol{F}(\boldsymbol{V})=Q \partial_x \boldsymbol{V}-\boldsymbol{P}.
\end{align} 
The cell averages at the next timestep are determined in the usual way by calculating the flux over the cell interfaces using a Lax-Friedrichs flux based on the implementation in \cite{KowalskiTorrilhon_shallowmomentapprox}. 

\subsection{The pressure update \texorpdfstring{$S_1$}{S1}}
This step computes $S_1(\delta_t)S_2(\delta_t)\boldsymbol{V}^n$ subsequent to the computation of $S_2(\delta_t)\boldsymbol{V}^n$.
 The general continuous Poisson-like equation as part of the LDSM system \eqref{eq:lin-moments-poisson} or non-linear DSM system \eqref{eq:general_DSM} takes the form 
\begin{align}
\label{eq:general-constraint}
    G\left(\boldsymbol{V},\partial_x\boldsymbol{V},\partial_{xx} \boldsymbol{V},h_b,\partial_xh_b,\partial_{xx} h_b\right)=0.
\end{align}
The function $G$ is linear in the highest derivatives $\partial_{xx} \boldsymbol{V}$, making it a quasi-linear PDE. Powers in the pressure variables $q_m$ and $\kappa_i$ or their derivatives are absent, transforming it into a linear PDE in the time-discrete case after the splitting \eqref{eq:solution-split}.
Therefore the solution $S_1(\delta_t)S_2(\delta_t)\boldsymbol{V}^n$ can be determined using a finite difference approximation. 

\subsubsection{Details on the finite difference discretization}

The mid-cell values of the transported variables are found with a zeroth-order reconstruction from the cell averages by just identifying averages and mid-cell values.

Central higher order finite differences are used on the grid of $n_x$ cell midpoints. For each variable $v \in \{h,u_m,\alpha_1,\dots,w_m,\gamma_1,\dots,q_m,\kappa_1,\dots\}$, either a second order, or a fourth order stencil can be used to replace the first and second order derivatives. The second order stencils read

\begin{align}
    \partial_{x} v_k = -\frac{v_{k-1} + v_{k+1}}{2 \Delta_x}, \qquad  
    \partial_{xx} v_k = \frac{v_{k-1}-{2 v_k}+{v_{k+1}}}{\Delta_x},
\end{align}
whereas the fourth order discretization is
\begin{align}
    \partial_{x} v_k &= \frac{v_{k-2}+{8 v_{k+1}}-{8 v_{k-1}}-{v_{k+2}}}{12 \Delta_x}, \\
    \partial_{xx} v_k &= \frac{-v_{k-2}+16 v_{k-1}-30 v_k+16 v_{k+1}-v_{k+2}}{12 \Delta_x^2}
\end{align}
where $\Delta_x={l_0}/{n_x}$ is the cell diameter.
The domain is assumed to be of length $l_0$, e.g. \(\left[-l_0/2,l_0/2\right]\).
For the experiments of this paper, the fourth order stencil was chosen. As far as is known, moving to simpler second order coefficients does not effect the results.

On a periodic boundary, the following identities hold:
\begin{align}
v_{n+1}=v_{1}, \qquad
v_{n+2}=v_{2}, \qquad
v_0 = v_{n}, \qquad
v_{-1}=v_{n-1}
\end{align}
The discretization is applied to the constraint equation \eqref{eq:general-constraint}. The emerging linear system takes the form  
\begin{equation}
\label{eq:fd-system}
    (I_{n \times n}-A) \boldsymbol{P} = \boldsymbol{S}
\end{equation} with the vector of unknown pressure variables
\begin{align}
    \boldsymbol{P}=(q_{m,1},\kappa_{1,1},\dots,\kappa_{N,1},q_{m,2},\kappa_{1,2},\dots,\kappa_{N,2},\dots,q_{m,n_x},\kappa_{1,n_x},\dots,\kappa_{N,n_x}).
\end{align} For the LDSM systems, the matrix A can conveniently be stated in explicit form (whereas for the DSM systems very large terms emerge):
\begin{equation}
   A=
   \begin{pmatrix}
   -\frac{5 h_0^2}{2 \Delta_x^2} & \frac{4 h_0^2}{3 \Delta_x^2} & -\frac{h_0^2}{12 \Delta_x^2} &  \mathbf{0} & -\frac{h_0^2}{12 \Delta_x^2} & \frac{4 h_0^2}{3 \Delta_x^2} \\
 \frac{4 h_0^2}{3 \Delta_x^2} & -\frac{5 h_0^2}{2 \Delta_x^2} & \frac{4 h_0^2}{3 \Delta_x^2} & -\frac{h_0^2}{12 \Delta_x^2} & \mathbf{ 0} & -\frac{h_0^2}{12 \Delta_x^2} \\
 -\frac{h_0^2}{12 \Delta_x^2} & \frac{4 h_0^2}{3 \Delta_x^2} & -\frac{5 h_0^2}{2 \Delta_x^2} & \frac{4 h_0^2}{3 \Delta_x^2} & -\frac{h_0^2}{12 \Delta_x^2} & \mathbf{0} \\
 \vdots& \vdots& \vdots & \vdots & \vdots & \vdots
 \\
 \mathbf{0} & -\frac{h_0^2}{12 \Delta_x^2} & \frac{4 h_0^2}{3 \Delta_x^2} & -\frac{5 h_0^2}{2 \Delta_x^2} & \frac{4 h_0^2}{3 \Delta_x^2} & -\frac{h_0^2}{12 \Delta_x^2} \\
 -\frac{h_0^2}{12 \Delta_x^2} & \mathbf{0} & -\frac{h_0^2}{12 \Delta_x^2} & \frac{4 h_0^2}{3 \Delta_x^2} & -\frac{5 h_0^2}{2 \Delta_x^2} & \frac{4 h_0^2}{3 \Delta_x^2} \\
  \frac{4 h_0^2}{3 \Delta_x^2} & -\frac{h_0^2}{12 \Delta_x^2} & \mathbf{0} & -\frac{h_0^2}{12 \Delta_x^2} & \frac{4 h_0^2}{3 \Delta_x^2} & -\frac{5 h_0^2}{2 \Delta_x^2}
   \end{pmatrix}.
\end{equation}
The matrix shown is valid for the fourth-order stencil and the corresponding (fourth-order stencil, LDSM) right hand side $S$ is given by 
\begin{align}
s_k=g h_k +g h_{b,k}+\frac{-h_{b,k-2}+16 h_{b,k-1}-30 h_{b,k}+16 h_{b,k+1}-h_{b,k+2}}{12 \Delta_x^2}. 
\end{align}
Note that although the matrix $A$ is singular due to the periodic boundary, the system \eqref{eq:fd-system} is usually uniquely solvable. For the actual system matrix $(I_{n \times n}-A)$ we have for \(\boldsymbol{v}\neq 0\)
\begin{equation}
    (I-A)\boldsymbol{v}=0 \quad\Leftrightarrow\quad (A\boldsymbol{v}=\boldsymbol{v})\quad\Leftrightarrow\quad \lambda=1 \text{ is eigenvalue of }A.
\end{equation} The eigenvalues of A depend on the configuration of $h_0$ and $\Delta_x$, so the matrix can become singular in theory but equation \eqref{eq:fd-system} is well posed in general.

The system is solved using a standard linear algebra routine.
\subsubsection{Second order method}
The splitting technique \eqref{eq:solution-split} is integrated into the finite volume method used in \cite{Cada2009,Schmidtmann2016,KowalskiTorrilhon_shallowmomentapprox}. More precisely, \eqref{eq:time-step} with \eqref{eq:solution-split} is computed in every stage of a three stage Runge-Kutta-Scheme, yielding a second order method.

\subsection{Results}

\begin{figure}[t]
    \centering
    \includegraphics[width=\linewidth]{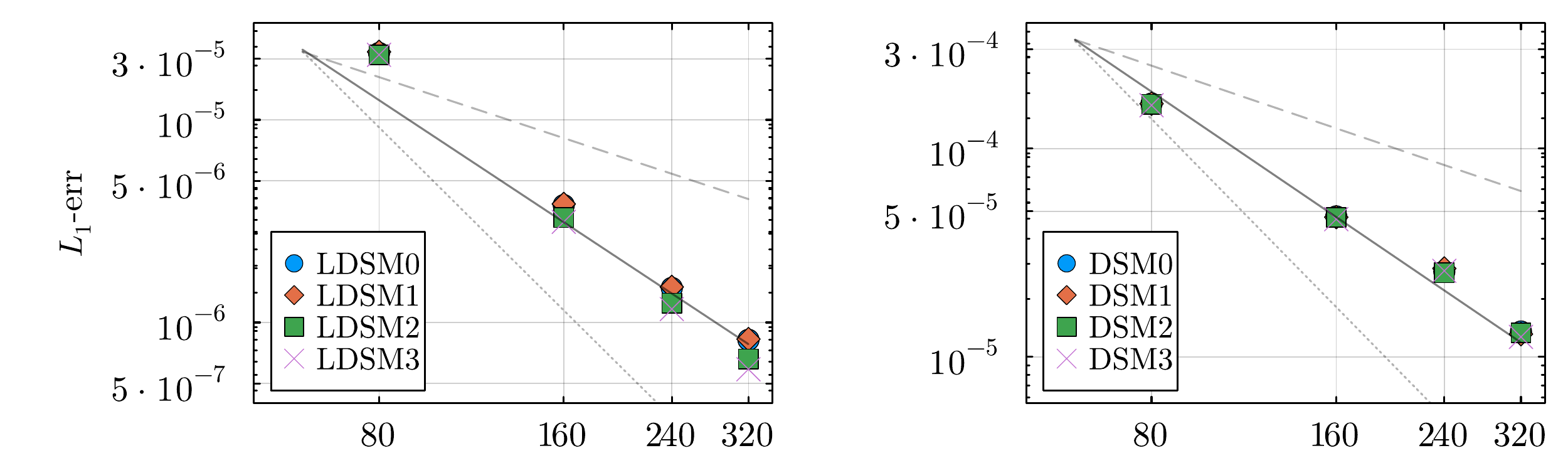}
    \caption{Error $err_{model}=\frac{1}{n_x}\sum_{k=1}^{n_x}||h_{manf.}(x_k)-h_{model}(x_k)||$ measured as average point-wise distance of the fluid height $h$ compared to the manufactured solution at time $t=0.4$; $CFL=0.9$. The plots (left: linear, right: nonlinear) show second order convergence of the method with respect to the grid resolution $n_x$.}
    \label{fig:manufactured}
\end{figure}

The code for all experiments is publicly available on GitHub \cite{Github_splitting_scheme}. The method was successfully tested against a manufactured solution simulating a traveling sine wave showing the expected second order convergence with respect to the resolution $n_x$ (see figure \ref{fig:manufactured}). A spacial resolution of $n_x>200$ was found to be sufficiently high to resolve any dispersive effects at feasible computational costs. 
The goal of the following numerical experiments is to show accuracy and robustness of the non-stationary DSM equations and the projection method. Further, differences to the existing shallow water moment equations should be discussed.
The numerical section of this paper treats two different experiments, both on a periodic domain of length $l_0=2$ with a typical water height $h_0$ and a background velocity $u_0$. Typical height and background velocity are not fixed which is reflected in the dimensionless experiments through varying values for the shallowness parameter and Froude number. The first experiment considers a flow over uneven bottom topography. The water surface is initially flat. 
The second experiment considers a traveling wave. It is taken from
\cite{KowalskiTorrilhon_shallowmomentapprox} where the initial condition for the height is $h(x,0)=h_0+h_0 e^{ 3 \cos \left(\pi  \left(x+\frac{1}{2}\right)\right)-4}$, a smooth symmetric wave.

\subsubsection{Flow over uneven bottom topography: The linear LDSM systems}

\begin{figure}[t!]
    \centering
    \includegraphics[width=.85\linewidth]{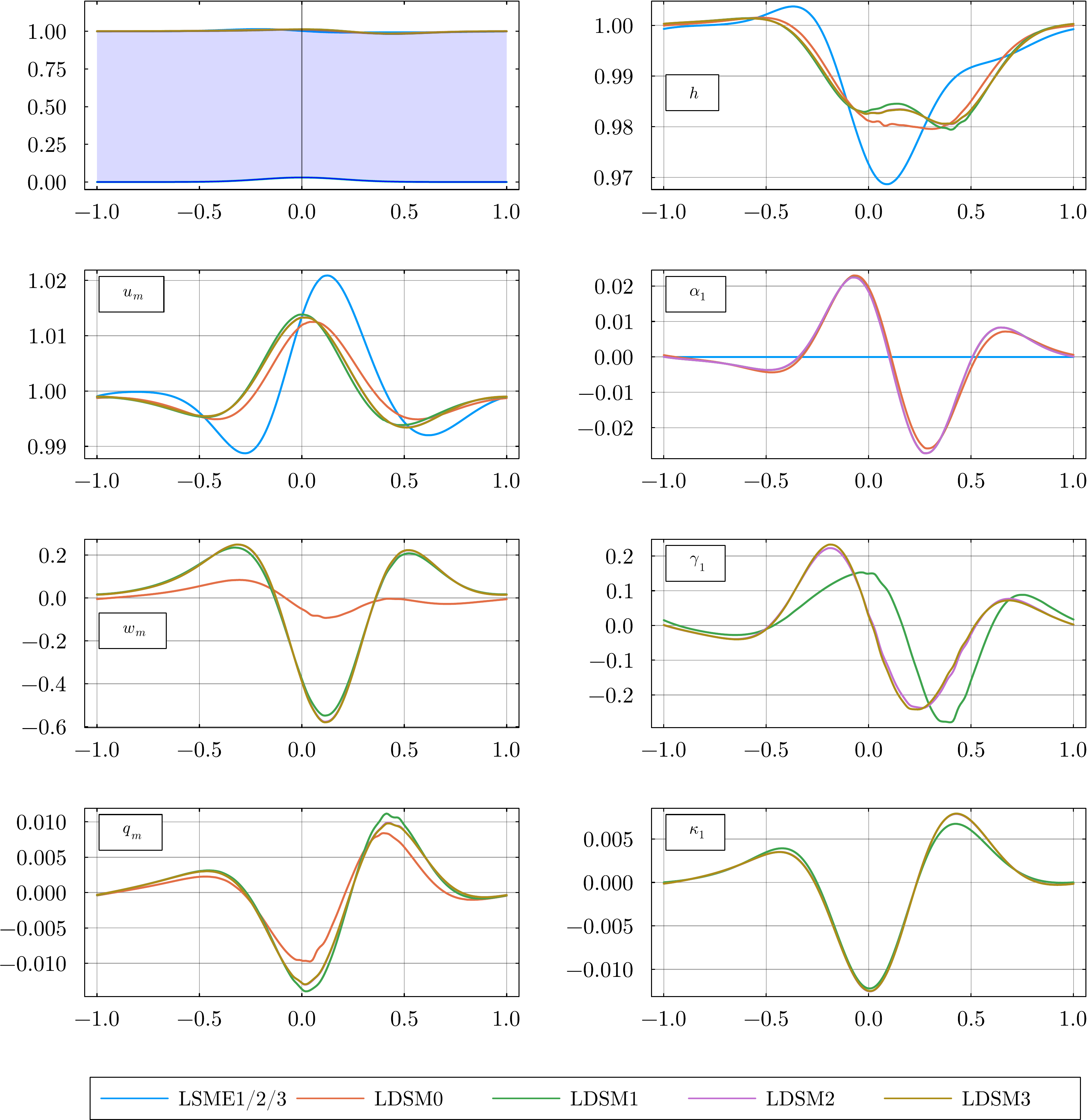}
    \caption{Experiment with hill-shaped bottom topography performed with LDSM0 (orange), LDSM1 (green), LDSM2 (purple), LDSM3 (brown), LSME1/2/3 (blue) at $\He=6/100, \Fr=3/4$. Depicted are (left-to-right, top-to-bottom) a view of the setup with actual water height $h+h_b$, the height h, mean of the horizontal velocity $u_m$ and the first moment $\alpha_1$, mean of the vertical velocity $w_m$ and first moment $\gamma_1$, non-hydrostatic pressure mean $q_m$ and first moment $\kappa_1$. The results obtained with the dispersive models deviate from the SME solution.}
    \label{fig:bottom-topography}
\end{figure}
The first experiment considers a flow over uneven bottom topography. For this purpose the bottom~$h_b$ was set to be hill-shaped with a maximal elevation of 3\% of the typical flow height: $h_b(x)=2 \He \eta \exp \left(-\frac{x^2}{2 \sigma ^2}\right)$ with $\eta=3/100$ and $\sigma=1/5$ at a degree of shallowness $\He=6/100$ (see top left plot of figure~\ref{fig:bottom-topography}).
Unlike in the previous experiments, the initial surface was set to be completely flat: $h(x,0)+h_b(x)=h_0$. However, there is a fast background flow of $\Fr=3/4$. Due to the proximity to equilibrium $h\approx h_0$, $u \approx u_0$ the use of linearized equations is justified. The experiment is run with the dispersive LDSM1-3 systems and the linearized classical shallow moment systems SME1-3.

The outcome is presented in figure~\ref{fig:bottom-topography}. The results of the LSME systems are indistinguishable and therefore combined. The LDSM results clearly differ from the LSME results in the fact that due to wave dispersion the fluctuation in water height is less pronounced. Also the maximum speed is reached directly above the summit rather than after it. There are visible differences within the LDSM systems with DSM0 possibly showing a deviation towards the SME solutions due to the lower non-hydrostatic pressure. 

This experiment shows that even close to equilibrium and with only minor bottom topography, dispersive effects play a role. As the deviation of DSM0 from the other systems shows, modeling higher order moments can be reasonable in this context as well.

\subsubsection{Flow over uneven bottom topography: Where the linear equations become inaccurate}

\begin{figure}[t!]
    \centering
    \includegraphics[width=0.85\linewidth]{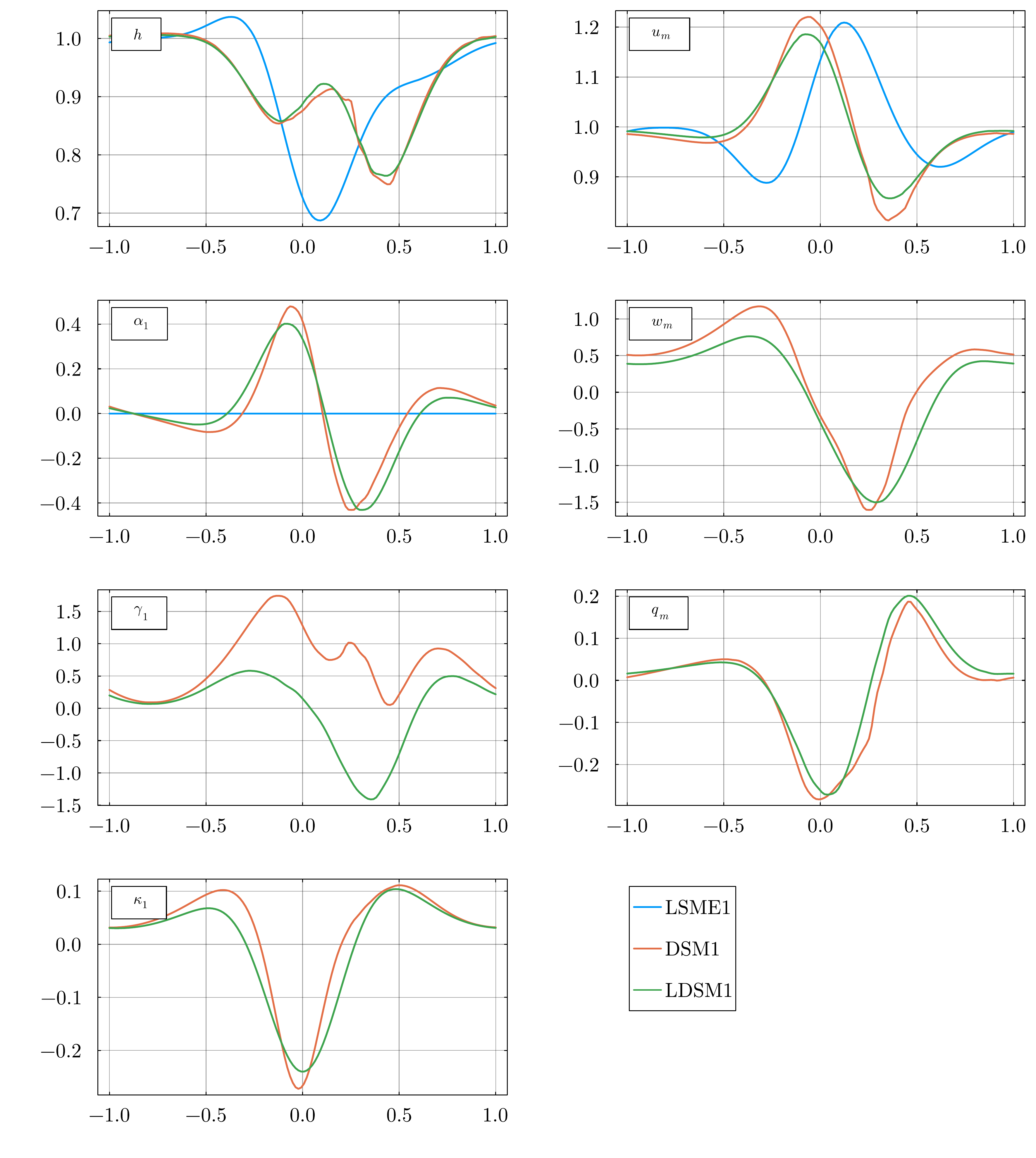}
    \caption{Results of the experiment with uneven bottom topography in a less shallow setting ($\He=18/100$). The linear and nonlinear results show significant differences; including a shock wave at $x=0.02$ for the nonlinear result. The linear LSME1 solution is far away from both dispersive solutions.}
    \label{fig:linear-vs-nonlinear}
\end{figure}
In the previous experiment results obtained with the linear and the nonlinear DSM equations were indistinguishable. The reason was the flow's proximity to equilibrium state. For other parameter settings, however, the solutions start to diverge. 
Consider a similar experiment with the following parameters: A flow with Froude number $\Fr=3/4$ over bottom topography $h_b(x)=2 \He \eta \exp \left(-\frac{x^2}{2 \sigma ^2}\right)$ with $\eta=3/100$ and $\sigma=1/5$ as above, but in a less shallow setting where $\He=18/100$.

Figure \ref{fig:linear-vs-nonlinear} shows results obtained with the DSM1, the LDSM1 and the LSME1 system. The solutions are similar to those of the previous experiment; for example, there is less variation in height for the non-dispersive equations, and the maximum of $u_m$ is shifted. Again, there is no first moment of the horizontal velocity for LSME. However, differences emerge between the linear and nonlinear solutions. A shock in the height can be observed at $x = 0.2$. The nonlinear solution is generally steeper and the extremes are more pronounced in plots of $u_m$ and $\alpha_1$. For the vertical velocity, $\gamma_1$ is estimated very differently by the linear and the nonlinear equations. Precisely, the profiles for the vertical velocity are facing into opposite directions around $x=0.2$.

\subsubsection{Traveling wave experiment: Effect of the shallowness parameter S}

The dispersion relation of the DSM equations yields high dispersivity for large $\He$ and convergence to the shallow water solution in the limit $\He \rightarrow 0$ \cite{ScholzKowalskiTorrilhon_dispersionshallowmoment}. This result is confirmed by the outcome of the second experiment without bottom topography and without background flow ($\Fr=0$) which was performed for different degrees of shallowness. 
\begin{figure}[t]
    \includegraphics[width=0.9\linewidth]{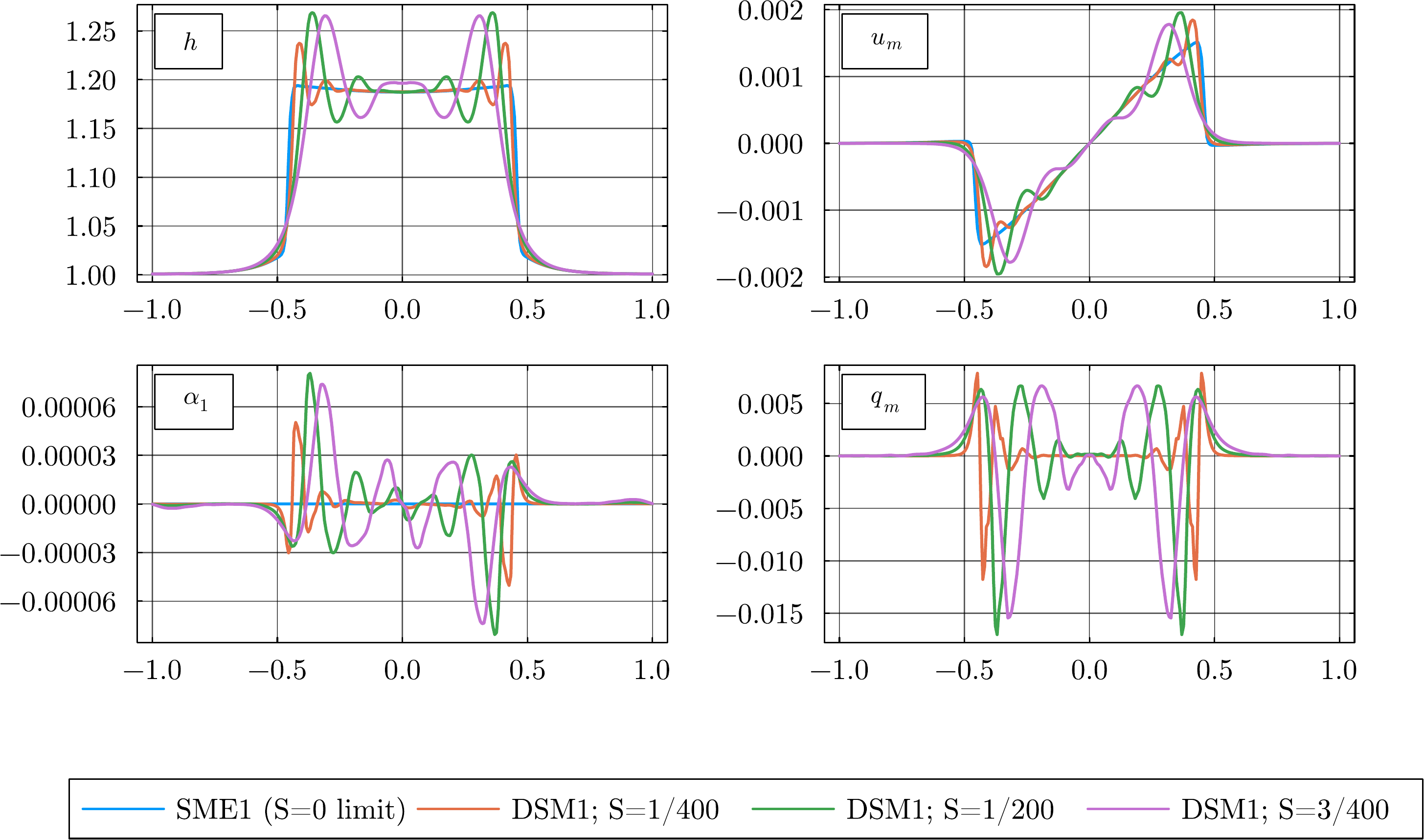}
    \caption{Results of the traveling wave without background flow (Fr=0) for the shallow water limit (blue) and $\He=1/400$ (orange), $\He=2/400$ (green), $\He=3/400$ (purple). Depicted are (left-to-right, top-to-bottom) the height h, mean of the horizontal velocity $u_m$, first moment of the horizontal velocity $\alpha_1$ and non-hydrostatic pressure $q_m$. The dispersive models exhibit the typical oscillating pattern and diverge from the shallow water solution as $\He$ increases.}
    \label{fig:convergenceH}
\end{figure}
The plots for height $h$ and velocity $u_m$ in figure~\ref{fig:convergenceH} show how the initially uniform wave becomes more and more dispersed at $t_{end}=2/\sqrt{g h_0}$ as $\He$ is increased. The dispersive effect can be attributed to the non-hydrostatic pressure component $q_m$. It also creates a horizontal velocity gradient visible in the plot of the first moment $\alpha_1$ of the horizontal velocity $u$.

\begin{figure}[h!]
    \centering
    \includegraphics[width=\linewidth]{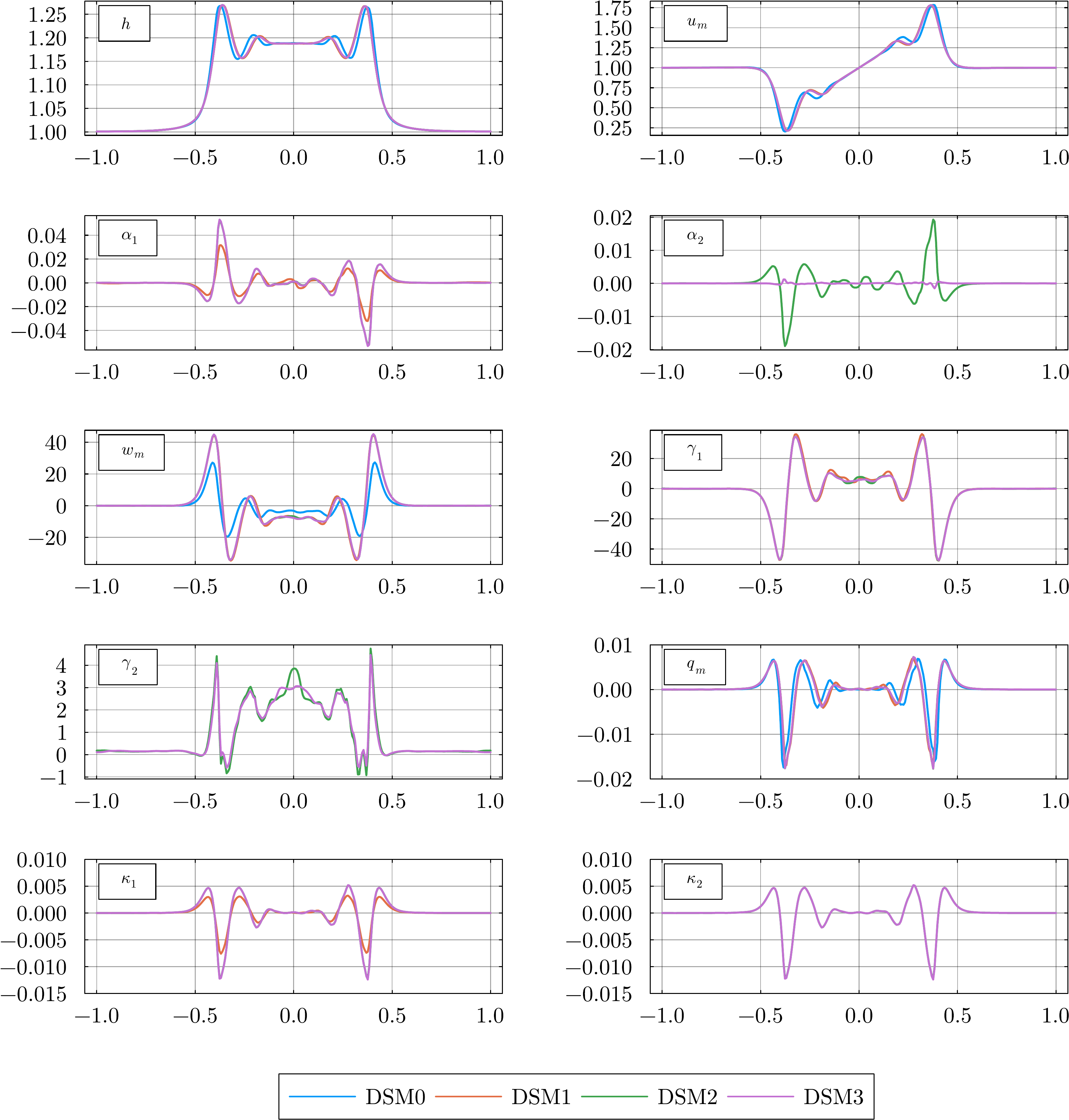}
    \caption{Traveling wave experiment with $\He=1/200$ and $\Fr=1/4$ for different systems of the hierarchy: DSM0 (blue), DSM1 (orange), DSM2 (green), DSM3 (purple). Depicted are (left-to-right, top-to-bottom) the mean of the horizontal velocity $u_m$, the moments of the horizontal velocity $\alpha_1$ and $\alpha_2$, the mean of the vertical velocity $w_m$, the corresponding moments $\gamma_1$ and $\gamma_2$, the non-hydrostatic pressure $q_m$ with moments $\kappa_1$ and $\kappa_2$. The differences between the solutions decrease as the hierarchical level increases.}
    \label{fig:dsm-hierarchy}
\end{figure}

\subsubsection{Traveling wave experiment: Accuracy for different DSM levels}

\begin{figure}[t!]
\centering
\includegraphics[width=0.45\linewidth]{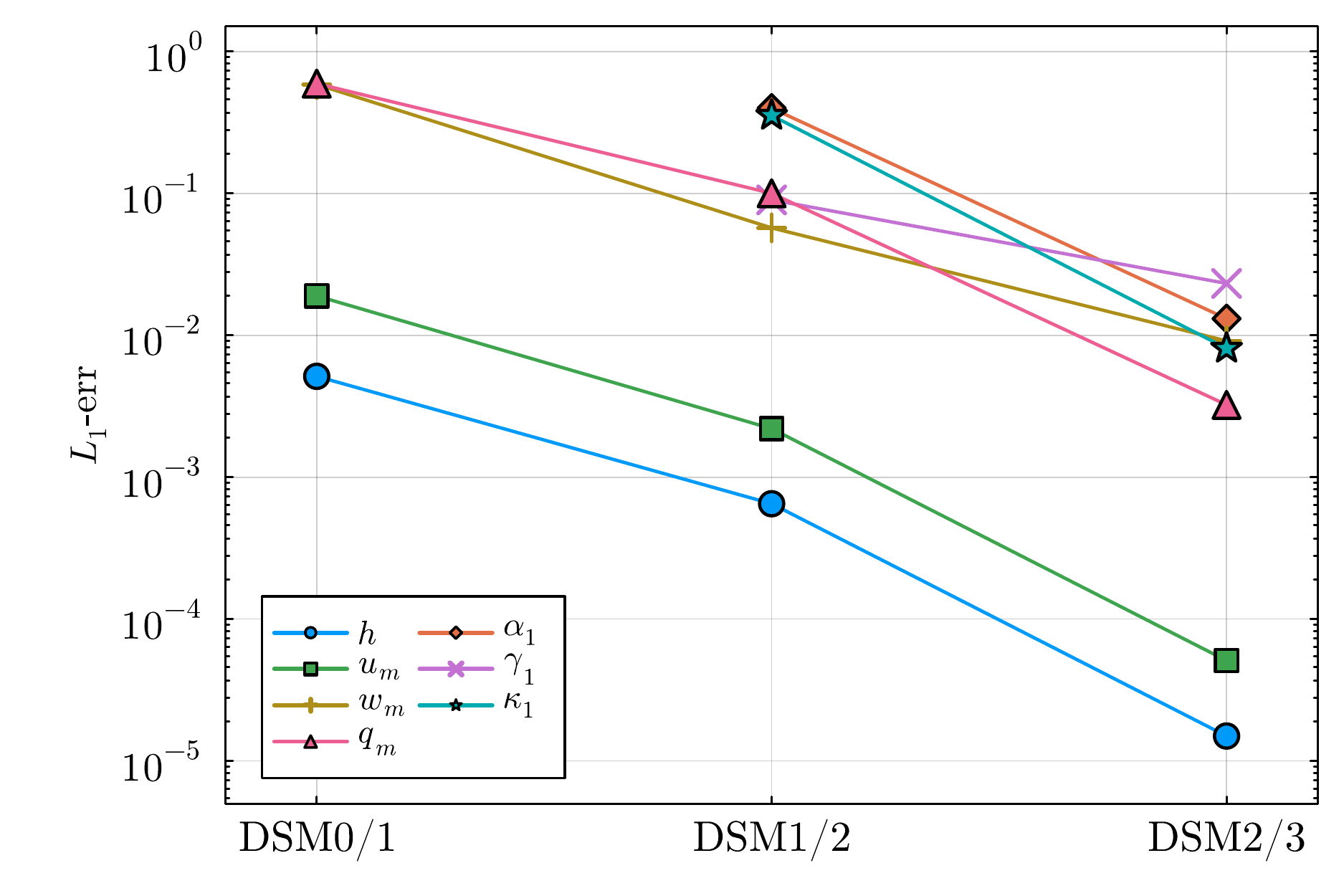}
\caption{L1-Differences between DSM0/1, DSM1/2 and DSM2/3 in the different flow variables. The differences between the solutions decrease visibly as the hierarchical level increases. The models are apparently approaching the (unknown) solution of the vertically resolved system.}
\label{fig:dsm-hierarchy-err} 
\end{figure}

Since the number of equations and thus the information contained increases with going up the DSM hierarchy, more accurate results for the higher-level systems can be expected. This was confirmed by the standard experiment with background flow ($\Fr=0.25$) which was run with the systems DSM0 to DSM3 at a degree of shallowness of $\He=1/200$. The result for the height $h$ is depicted in the upper left plot in figure $\ref{fig:dsm-hierarchy}$ and for the other flow variables it is displayed in the other nine plots of figure~\ref{fig:dsm-hierarchy}. In the means $u_m$, $w_m$ and $q_m$, plotted in figure~\ref{fig:dsm-hierarchy}, results for DSM1 to DSM3 are indistinguishable. In the first moments, however, a deviation can still be seen for DSM1. Finally, the second moments exhibit that DSM2 and DSM3 do not actually coincide. The models are apparently approaching the (unknown) solution of the vertically resolved system. The convergence plot in figure~\ref{fig:dsm-hierarchy-err} captures the observed model convergence by displaying the L1-difference from one system to the next highest in the hierarchy for each flow variable for the first three DSM systems. It could be described as a hierarchical convergence since we do not compare to the true solution and consider the moments separately. However, we see that the hierarchical convergence is faster for the mean values than for the higher moments and it is present in all moments such that not only their linear combination is converging. We expect this effect to carry over to higher-order moments, so first order moments should converge faster than the second order ones.

\section{Conclusion}\label{sec:conclusion}

This paper introduced a splitting method for solving the nonstationary linear and nonlinear Dispersive Shallow Moment Equations. The derivation of the models was presented in a new notation. It was shown for the reference systems as well as for specific moment systems how a Poisson-like system can be derived that allows for a de-coupling of the advected variables from the pressure variables in the time-discrete setting.
Numerical simulations confirm that the introduced splitting method is stable and produces physically reasonable solutions; while the results on dispersivity of the DSM equations had been more theoretical in nature up to now, these effects could now be observed directly. It leads to the conclusion that the DSM models are suitable equation systems for flows of intermediate depth where non-hydrostatic effects play a role and the vertical profile of the horizontal velocities and the pressure is nontrivial.
Possible future work should advance the DSM equation further toward practical applicability. This includes experiments using a non-periodic domain, which raises questions on the boundary conditions.


\section*{Acknowledgments}

This work is partially funded by the Deutsche Forschungsgemeinschaft (DFG, German Research Foundation) – 320021702/GRK2326 – Energy, Entropy, and Dissipative Dynamics (EDDy).


\appendix

\section{Appendix}

\subsection{Reformulated Reference System}\label{appx:subsec:equiv_ref_sys}

We show here more details on the equivalence of the two systems~\eqref{eq:free-surfcae-Euler-ref-mapped} together with~\eqref{eq:mapped-ref-sys-boundary} and~\eqref{eq:free-surface-Euler-comp}. For the notation compare to section~\ref{subsec:mapped_ref_sys}.\\
First, we have for an arbitrary function \(\psi(t,x,y,z)\) and its mapped counterpart \(\tilde{\psi}(t,x,y,\zeta)\) the relations
\begin{subequations}\label{eq:mapped-deri-rel}
	\begin{equation}
		\label{eq:mapped_deri_rel_hori}
		h\partial_s\psi = \partial_s\left(h\tilde{\psi}\right) -\partial_{\zeta}\left(\partial_s(\zeta h + h_b)\tilde{\psi}\right),
	\end{equation}
	for \(s\in\{t,x,y\}\) and for the remaining derivative
	\begin{equation}
		\label{eq:mapped_deri_rel_vert}
		h\partial_z\psi = \partial_{\zeta}\tilde{\psi}.
	\end{equation}
\end{subequations}
We expand equations~\eqref{eq:free-surfcae-Euler-ref-mapped}, multiply by \(h\) and use the relations above. This yields for equation \eqref{eq:free_surface_Euler_ref_divergence_mapped}, neglecting the tilde for readability, 
\begin{equation}
	\label{eq:free_surface_Euler_ref_divergence_mapped_expanded}
	\partial_x(hu) + \partial_y(hv) + \partial_{\zeta}w - \partial_{\zeta}(\partial_x(\zeta h + h_b)u) - \partial_{\zeta}(\partial_y(\zeta h + h_b)v) = 0,
\end{equation}
what is obviously equal to~\eqref{eq:free_surface_Euler_divergence}.\\
For equation~\eqref{eq:free_surface_Euler_ref_evolution_mapped} we can now use this equation too. This gives us with the relations~\eqref{eq:mapped-deri-rel}
\begin{equation}
	\label{eq:free_surface_Euler_ref_evolution_mapped_expanded}
	\begin{split}
		&\partial_t(h\boldsymbol{u}) + \partial_x(hu\boldsymbol{u}) + \partial_y(hv\boldsymbol{u}) \\ 
		&+ \partial_{\zeta}\left(\left(w - \partial_t(\zeta h + h_b)
		-\partial_x(\zeta h + h_b)u - \partial_y(\zeta h + h_b) v\right) \boldsymbol{u}\right) = -h\left((J_{\Sigma})^{\rm T}\tilde{D}_x\right)p + h\boldsymbol{g}.
	\end{split}
\end{equation}
If we reformulate the pressure term with~\eqref{eq:mapped-deri-rel} this equation is almost of the same form as~\eqref{eq:free-surface-Euler-comp}. We have to consider the \(\zeta\)-derivative in more detail.\\
Therefore, we integrate~\eqref{eq:free_surface_Euler_ref_divergence_mapped_expanded} and use the boundary condition at the bottom~\eqref{eq:mapped_ref_sys_bottom} to get an explicit expression for the vertical velocity \(w\)
\begin{equation}
	\label{eq:mapped_ref_sys_explicit_vert_velo}
	w = - \partial_x\left( h\int_0^{\zeta} u \,\mathrm{d}\hat{\zeta}\right) - \partial_y\left(h\int_0^{\zeta} v\,\mathrm{d}\hat{\zeta}\right) + u\partial_x(\zeta h + h_b) + y\partial_y(\zeta h + h_b).
\end{equation}
We use the boundary conditions~\eqref{eq:mapped-ref-sys-boundary} and take the difference of the vertical velocity at the surface and bottom, i.e., we calculate
\begin{equation}
	\label{eq:approach_height_balance}
	w_{|_{\zeta =1}} - w_{|_{\zeta = 0}}
\end{equation}
using the definition~\eqref{eq:mapped_ref_sys_explicit_vert_velo} of the vertical velocity \(w\).
This gives us the height balance~\eqref{eq:free_surface_Euler_height},
\begin{equation}
	\label{eq:height_balance}
	\partial_th + \partial_x(hu_m) + \partial_y(hv_m) = 0.
\end{equation}
Inserting~\eqref{eq:mapped_ref_sys_explicit_vert_velo} and the height balance~\eqref{eq:height_balance} into~\eqref{eq:free_surface_Euler_ref_evolution_mapped_expanded} we achieve
\begin{equation}
	\label{eq:mapped_ref_sys_rewritten}
	\begin{split}
		&\partial_t(h\boldsymbol{u}) + \partial_x(hu\boldsymbol{u}) + \partial_y(hv\boldsymbol{u}) \\
		&+ \partial_{\zeta}\left(\left( -\partial_x\left(h\int_0^{\zeta} u\,\mathrm{d}\hat{\zeta}\right) - \partial_y\left(h\int_0^{\zeta} v \,\mathrm{d}\hat{\zeta}\right) + \zeta (\partial_x(hu_m) + \partial_y(hv_m)\right)\boldsymbol{u}\right) = -\left((J_{\Sigma})^{\rm T}\tilde{D}_x\right)p + h\boldsymbol{g}.
	\end{split}
\end{equation}
According to~\cite{KowalskiTorrilhon_shallowmomentapprox} we can define the vertical coupling operator through
\begin{equation}
	\label{eq:vertical_coupling_operator}
	h\omega\left[h,u,v\right] = -\partial_x\left(h\int_0^{\zeta}(u-u_m)\,\mathrm{d}\hat{\zeta}\right) - \partial_y\left(h\int_0^{\zeta}(v-v_m)\,\mathrm{d}\hat{\zeta}\right).
\end{equation}
Comparing this definition to the \(\zeta\)-derivative of~\eqref{eq:mapped_ref_sys_rewritten} we can replace the term in the inner brackets with the vertical coupling operator
\begin{equation}
	\label{eq:mapped_ref_sys_final_form}
	\partial_t(h\boldsymbol{u}) + \partial_x(hu\boldsymbol{u}) + \partial_y(hv\boldsymbol{u}) 
	+ \partial_{\zeta}(h\omega\boldsymbol{u}) = -\left((J_{\Sigma})^{\rm T}\tilde{D}_x\right)p + h\boldsymbol{g}.
\end{equation}
From this point it is again an easy application of~\eqref{eq:mapped-deri-rel} to reformulate the pressure term and achieve equation~\eqref{eq:free_surface_Euler_evolution}.

\subsection{Higher Order Examples of Shallow Moment Systems}\label{appx:subsec:higher_order_examples}

\subsubsection{LDSM3}\label{appx:subsubsec:LDSM3}

\begin{subequations}\label{LDSM3}
	\begin{align}
		\label{eq:LDSM3_evol}
		&\partial_t\begin{pmatrix}
			h \\ \Fr^2u_m \\ \Fr^2 \alpha_1 \\ \Fr^2\alpha_2 \\ \Fr^2\alpha_3\\ \He^2\Fr^2 w_m \\ \He^2\Fr^2 \gamma_1 \\ \He^2\Fr^2\gamma_2 \\ \He^2\Fr^2\gamma_3
		\end{pmatrix} + \partial_x\begin{pmatrix}
			h + u_m \\ \Fr^2u_m + h + q_m \\ \Fr^2 \alpha_1 + \kappa_1 \\ \Fr^2 \alpha_2 + \kappa_2 \\ \Fr^2 \alpha_3 + \kappa_3\\ \He^2\Fr^2 w_m \\ \He^2\Fr^2 \gamma_1 \\ \He^2 \Fr^2 \gamma_2 \\\He^2\Fr^2 \gamma_3
		\end{pmatrix}  =  \begin{pmatrix}
			0 \\ 0 \\ 0 \\ 0 \\ 0 \\ q_m + \kappa_1 + \kappa_2 + \kappa_3 \\ -3q_m + 3\kappa_1 + 3\kappa_2 + 3\kappa_3 \\ 5q_m - 5\kappa_1 + 5\kappa_2 + 5\kappa_3 \\ -7q_m + 7\kappa_1 - 7\kappa_2 + 7\kappa_3
		\end{pmatrix} - \frac{1}{\He}\partial_xh_b\begin{pmatrix}
			0 \\ 1 \\ 0 \\ 0 \\ 0 \\ 0 \\ 0 \\ 0 \\ 0
		\end{pmatrix}, \\
		\label{eq:LDSM3_constraint_zeroth}
		&\partial_xu_m = \frac{1}{\He}\partial_xh_b - w_m + \gamma_1 - \gamma_2 + \gamma_3,\\
		\label{eq:LDSM3_constraint_first}
		&\partial_x\alpha_1 = \frac{3}{\He}\partial_xh_b - 3w_m -3\gamma_1 + 3\gamma_2 - 3\gamma_3, \\
		\label{eq:LDSM3_constraint_second}
		&\partial_x\alpha_2 = \frac{5}{\He}\partial_xh_b - 5w_m -5\gamma_1 -5\gamma_2 +5\gamma_3,\\
		\label{eq:LDSM3_constraint_third}
		&\partial_x\alpha_3 = \frac{7}{\He}\partial_xh_b - 7w_m -7\gamma_1 -7\gamma_2 -7\gamma_3.
	\end{align}
	In this case the Poisson equations for the averaged and higher order pressure moments are given by
	\begin{align}
		\label{eq:LDSM3_poisson_zeroth}
		\partial_{xx}q_m &= \frac{16}{\He^2}q_m - \frac{14}{\He^2}\kappa_1 + \frac{10}{\He^2}\kappa_2 - \frac{4}{\He^2}\kappa_3 -\partial_{xx}\left(h + \frac{1}{\He}h_b\right) - \frac{\Fr^2}{\He}\partial_{xx}h_b,\\
		\label{eq:LDSM3_poisson_first}
		\partial_{xx}\kappa_1 &= -\frac{42}{\He^2}q_m + \frac{48}{\He^2}\kappa_1 - \frac{24}{\He^2}\kappa_2 +\frac{18}{\He^2}\kappa_3 - \frac{3\Fr^2}{\He}\partial_{xx}h_b,\\
		\label{eq:LDSM3_poisson_second}
		\partial_{xx}\kappa_2 &= \frac{50}{\He^2}q_m - \frac{40}{\He^2}\kappa_1 +\frac{80}{\He^2}\kappa_2 + \frac{10}{\He^2}\kappa_3 - \frac{5\Fr^2}{\He}\partial_{xx}h_b,\\
		\label{eq:LDSM3_poisson_third}
		\partial_{xx}\kappa_3 &= -\frac{28}{\He^2}q_m + \frac{42}{\He^2}\kappa_1 + \frac{14}{\He^2}\kappa_2 +\frac{112}{\He^2}\kappa_3 - \frac{7\Fr^2}{\He}\partial_{xx}h_b.
	\end{align}
\end{subequations}

\subsubsection{Second order or quadratic system}\label{appx:subsubsec:DSM2}

\begin{subequations}\label{example:dsm2}
	\begin{equation}
		\label{eq:dsm2}
		\dt
		\begin{pmatrix}
			h \\
			\Fr^2 h u_m \\
			\Fr^2 h \alpha_1 \\
			\Fr^2 h \alpha_2 \\
			\He^2 \Fr^2 h w_m \\
			\He^2 \Fr^2 h \gamma_1 \\
			\He^2 \Fr^2 h \gamma_2 \\
		\end{pmatrix}
		+
		\dx
		\begin{pmatrix}
			h u_m \\
			\frac{h^2}{2}+\Fr^2(h u_m^2 +\frac{\alpha_1^2}{3}+\frac{\alpha_2^2}{2}) +h q_m \\
			\Fr^2 h (2 u_m\alpha_1 +\frac{4 \alpha_1 \alpha_2}{5}) +h \kappa_1\\
			\Fr^2 h (\frac{2\alpha_1^2}{3}+2 u_m\alpha_2 +\frac{2 \alpha_2^2}{7}) +h \kappa_1\\
			\He^2 \Fr^2 h( u_m w_m+ \frac{\alpha_1 \gamma_1}{3}+\frac{\alpha_2 \gamma_2}{5}) \\
			\He^2 \Fr^2 h (\alpha_1 w_m +  u_m \gamma_1+\frac{2 \alpha_2 \gamma_1}{5}+\frac{2 \alpha_1 \gamma_2}{5}) \\
			\He^2 \Fr^2 h (\alpha_2 w_m+  \frac{2 \alpha_1 \gamma_1}{3}+u_m \gamma_2+\frac{2 \alpha_2 \gamma_2}{7}) \\
		\end{pmatrix}
		=Q dx
		\begin{pmatrix}
			h\\
			\Fr^2 h u_m \\
			\Fr^2 h \alpha_1 \\
			\Fr^2 h \alpha_2 \\
			\He^2 \Fr^2 h w_m \\
			\He^2 \Fr^2h \gamma_1\\
			\He^2 \Fr^2h \gamma_2
		\end{pmatrix} -\boldsymbol{P}.
	\end{equation}
	Subject to the constraints
	\begin{align}
		& \dx u_m h + \dx h (\alpha_1-\alpha_2)+w_m -\gamma_1+\gamma_2 - \frac{1}{\He}(u_m -\alpha_1+\alpha_2) \dx h_b = 0, \hspace{20pt}& \\[1ex]
		& \dx \alpha_1 h - \dx h \alpha_1 +3 (w_m + \gamma_1-\gamma_2) - \frac{3}{\He}(u_m + \alpha_1+\alpha_2)\dx h_b = 0, \\[1ex]
		& \dx \alpha_2 h - \dx h (5\alpha_1-2\alpha_2) +5 (w_m + \gamma_1+\gamma_2) - \frac{5}{\He}(u_m - \alpha_1+\alpha_2)\dx h_b = 0,
	\end{align}
	with
	\begin{equation}
		\boldsymbol{P}=
		\begin{pmatrix}
			0 \\
			0 \\
			0 \\
			0 \\
			-q_m - \kappa_1-\kappa_2 \\
			3(q_m - \kappa_1-\kappa_2) \\
			-5(q_m - \kappa_1+\kappa_2) \\
		\end{pmatrix} +\frac{1}{\He}\dx h_b
		\begin{pmatrix}
			0 \\
			h+ q_m + \kappa_1+\kappa_2\\
			0 \\
			10 \kappa_1\\
			0 \\
			0 \\
			0
		\end{pmatrix}
	\end{equation}
	and
	\begin{equation}
		Q=
		\begin{pmatrix}
			0 &0 &0 &0 &0 &0 &0\\
			0 &0 &0 &0 &0 &0 &0\\
			2 \kappa_1 & 0 & \Fr^2 (u_m-\frac{\alpha_2}{5}) &\frac{\Fr^2 \alpha_1}{5} &0 &0 &0\\
			-5\kappa_1 +3 \kappa_2 & 0 & \Fr^2 \alpha_1 &\Fr^2 (u_m + \frac{\alpha_2}{7}) &0 &0 &0\\
			0 &0 &0 &0 &0 &0 &0 \\
			0 &0 & \He^2 \Fr^2 (w_m-\frac{\gamma_2}{5})& \frac{1}{5}\He^2 \Fr^2 \gamma_1 & 0 &0 &0\\
			0 &0 & \He^2 \Fr^2 \gamma_1& \He^2 \Fr^2 (w_m + \frac{\gamma_2}{7}) & 0 &0 &0
		\end{pmatrix}
	\end{equation}
\end{subequations}
for DSM2.

\subsubsection{Third order or cubic system}\label{appx:subsubsec:DSM3}

Finally we have the same generic structure
\begin{subequations}\label{example:dsm3}
	\begin{equation}
		\label{eq:dsm3}
		\dt \boldsymbol{V} + \dx \boldsymbol{F}(\boldsymbol{V}) = Q \dx \boldsymbol{V} - \boldsymbol{P},
	\end{equation}
	now with the variable vector
	
	\begin{equation}
		\boldsymbol{V}=\left(
		h,
		\Fr^2 h u_m,
		\Fr^2 h \alpha_1,
		\Fr^2 h \alpha_2,
		\Fr^2 h \alpha_3,
		\He^2 \Fr^2 h w_m,
		\He^2 \Fr^2 h \gamma_1,
		\He^2 \Fr^2 h \gamma_2,
		\He^2 \Fr^2 h \gamma_3 \right)
	\end{equation}
	and
	
	\begin{equation}
		\boldsymbol{F}(\boldsymbol{V})=
		\begin{pmatrix}
			h u_m \\
			\frac{h^2}{2}+\Fr^2(h u_m^2 +\frac{\alpha_1^2}{3}+\frac{\alpha_2^2}{2}+\frac{\alpha_3^2}{7}) +h q_m\\
			\Fr^2 h (2 u_m\alpha_1 +\frac{4 \alpha_1 \alpha_2}{5}+\frac{18 \alpha_2 \alpha_3}{35}) +h \kappa_1\\
			\Fr^2 h (\frac{2\alpha_1^2}{3}+2 u_m\alpha_2 +\frac{2 \alpha_2^2}{7}+\frac{6 \alpha_1 \alpha_3}{7}+\frac{4 \alpha_3^2}{21}) +h \kappa_1\\
			\Fr^2 h (\frac{6\alpha_1 \alpha_2^2}{5}+2 u_m\alpha_3 +\frac{8 \alpha_2\alpha_3}{15}) +h \kappa_1\\
			\He^2 \Fr^2 h( u_m w_m+ \frac{\alpha_1 \gamma_1}{3}+\frac{\alpha_2 \gamma_2}{5}+\frac{\alpha_3\gamma_3}{7}) \\
			\He^2 \Fr^2 h (\alpha_1 w_m +  u_m \gamma_1+\frac{2}{5}( \alpha_2 \gamma_1+ \alpha_1 \gamma_2)+\frac{9}{35}(\alpha_3 \gamma_3+\alpha_2 \gamma_3) \\
			\He^2 \Fr^2 h (\alpha_2 w_m +  \frac{2 \alpha_1 \gamma_1}{3}+\frac{3}{7}( \alpha_3 \gamma_1 + \alpha_1 \gamma_3)+u_m \gamma_2+\frac{2 \alpha_2 \gamma_2}{7}+\frac{4}{\alpha_3 \gamma_3}{21}) \\
			\He^2 \Fr^2 h (\alpha_3 w_m + \frac{3}{5} (\alpha_2 \gamma_1 + \alpha_1 \gamma_2) + u_m \gamma_3 + \frac{4}{15} (\alpha_3 \gamma_2 + \alpha_2 \gamma_3)
		\end{pmatrix}
	\end{equation}
	and constraints
	\begin{align}
		& \dx u_m h + \dx h (\alpha_1-\alpha_2+\alpha_3)+w_m -\gamma_1+\gamma_2-\gamma_3 - \frac{1}{\He} (u_m -\alpha_1+\alpha_2-\alpha_3) \dx h_b = 0, \\
		& \dx \alpha_1 h - \dx h \alpha_1 +3 (w_m + \gamma_1-\gamma_2+\gamma_3) - \frac{3}{\He}(u_m + \alpha_1+\alpha_2+\alpha_3)\dx h_b = 0, \\
		& \dx \alpha_2 h - \dx h (5\alpha_1-2\alpha_2) +5 (w_m + \gamma_1+\gamma_2-\gamma_3) - \frac{5}{\He}(u_m - \alpha_1+\alpha_2+\alpha_3)\dx h_b = 0,\\
		& \dx \alpha_3 h - \dx h (-7\alpha_1\!+\!7\alpha_2\!-\!3\alpha_3) +7 (w_m\! +\! \gamma_1\!+\!\gamma_2\!+\!\gamma_3) - \frac{7}{\He}(u_m\!+\!\alpha_1\!-\!\alpha_2\!+\!\alpha_3)\dx h_b = 0,
	\end{align}
	with
	\begin{equation}
		\boldsymbol{P}=
		\begin{pmatrix}
			0 \\
			0 \\
			0 \\
			0 \\
			0 \\
			-q_m - \kappa_1-\kappa_2 - \kappa_3\\
			3(q_m - \kappa_1-\kappa_2-\kappa_3) \\
			-5(q_m - \kappa_1+\kappa_2+\kappa_3) \\
			7(q_m - \kappa_1 + \kappa_2 - \kappa_3)
		\end{pmatrix} +\frac{1}{\He}\partial_x h_b
		\begin{pmatrix}
			0 \\
			h+ q_m + \kappa_1+\kappa_2+\kappa_3\\
			0 \\
			10 \kappa_1\\
			14 \kappa_2 \\
			0 \\
			0 \\
			0
		\end{pmatrix}
	\end{equation}
	and
	\begin{equation}
		Q=
		\begin{pmatrix}
			0 &0 &0 &0 &0 & \\
			0 &0 &0 &0 &0 & \\
			2 \kappa_1 & 0 & \Fr^2 (u_m-\frac{\alpha_2}{5}) &\Fr^2 \frac{ \alpha_1}{5}-\frac{3 \alpha_3}{35} &\frac{3 \Fr^2 \alpha_2}{35} &\\
			-5\kappa_1 +3 \kappa_2 & 0 & \Fr^2 (\alpha_1-\frac{3 \alpha_3}{7} &\Fr^2 (u_m + \frac{\alpha_2}{7}) &\Fr^2(\frac{2 \alpha_1}{7} + \frac{\alpha_3}{21}) &\\
			7 \kappa_1 - 7 \kappa_2 + 4 \kappa_3 & 0 & \frac{6 \Fr^2 \alpha_2}{5} & \Fr^2(\frac{4 \alpha_1}{5}+\frac{2 \alpha_3}{15}) & \Fr^2 (u_m + \frac{\alpha_2}{5}) & \boldsymbol{0}_{4\times9}\\
			0 &0 &0 &0 &0 & \\
			0 &0 & \He^2 \Fr^2 (w_m-\frac{\gamma_2}{5})& \He^2 \Fr^2 \frac{\gamma_1}{5}-\frac{3 \gamma_3}{35} & \frac{3}{35} \He^2 \Fr^2 \gamma_2 &\\
			0 &0 & \He^2 \Fr^2 (\gamma_1-\frac{3 \gamma_3}{7})& \He^2 \Fr^2 (w_m + \frac{\gamma_2}{7}) & \He^2 \Fr^2 (\frac{2 \gamma_1}{7}+\frac{\gamma_3}{21}) &\\
			0 &0 & \frac{6}{5} \He^2 \Fr^2 \gamma_2 & \He^2 \Fr^2 (\frac{4 \gamma_1}{5}+\frac{2 \gamma_3}{15} & \He^2 \Fr^2 (w_m + \frac{\gamma_2}{5}) &
		\end{pmatrix}
	\end{equation}
\end{subequations}
for DSM3.

\subsection{General Dispersive Shallow Moment System}\label{appx:subsec:DSM-N}

\subsubsection{Abbreviations}\label{appx:subsubsec:abbr}

For a shorter notation we define the following abbreviations
\begin{subequations}\label{appx:eq:abbreviations}
	\begin{align}
		\label{appx:eq:abbreviations_G}
		G_{ij} &= (2i+1)\int_0^1 \phi_i'\phi_j\,\mathrm{d}\zeta,\\
		\label{appx:eq:abbreviations_A}
		A_{ijk} &= (2i+1)\int_0^1\phi_i\phi_j\phi_k\,\mathrm{d}\zeta,\\
		\label{appx:eq:abbreviations_H}
		H_{ij} &= (2i+1)\int_0^1\zeta \phi_i'\phi_j\,\mathrm{d}\zeta, \\
		\label{appx:eq:abbreviations_B}
		B_{ijk} &= (2i+1)\int_0^1 \phi_i'\left(\int_0^\zeta \phi_j\,\mathrm{d}\hat{\zeta}\right)\phi_k\,\mathrm{d}\zeta.
	\end{align}
\end{subequations}
These are used to write the DSM systems in are more suitable way.

\subsubsection{Derivation DSM-N}\label{appx:subsubsec:deri_dsmN}

The reference system with non-hydrostatic pressure is given by~\eqref{eq:free-surface-Euler-comp} and reads after rewriting the pressure term
\begin{subequations}\label{appx:eq:ref-sys}
	\begin{align}
		\label{appx:eq:ref_sys_height}
		&\partial_th + \partial_x(hu_m) + \partial_y(hv_m) = 0,\\
		\label{appx:eq:ref_sys_constraint}
		&\partial_x(hu) + \partial_y(hv) + \partial_{\zeta}w - \partial_{\zeta}(\partial_x(\zeta h + h_b)u) - \partial_{\zeta}(\partial_y(\zeta h + h_b)v)=0,\\
		\label{appx:eq:ref_sys_velo}
		&\partial_t(h\boldsymbol{u}) + \partial_x(hu\boldsymbol{u})+\partial_y(hv\boldsymbol{u})+\partial_{\zeta}(h\omega\boldsymbol{u})+\frac{g}{2}\tilde{\nabla}_x\left(h^2\right)+hg\tilde{\nabla}_x h_b + \left((J_{\Sigma})^{\rm T}\tilde{D}\right)_x q =\boldsymbol{0},
	\end{align}
\end{subequations}
with the boundary conditions
\begin{equation}
	\label{appx:eq:boundaries}
	q_{|_{\zeta =1}}=0\qquad \text{and}\qquad w_{|_{\zeta =0 }}=\boldsymbol{u}_{|_{\zeta =0}}\cdot \tilde{\nabla}_x h_b
\end{equation}
and the vertical coupling operator, (compare~\eqref{eq:vertical_coupling_operator})
\begin{equation}
	\label{appx:eq:vertical_coupling_op}
	\omega\left[h,u,v\right] = - \frac{1}{h}\left(\partial_x\left(h\int_{0}^{\zeta}(u - u_m)\,\mathrm{d}\hat{\zeta}\right)+\partial_y\left(h\int_{0}^{\zeta}(v - v_m)\,\mathrm{d}\hat{\zeta}\right)\right).
\end{equation}
Here we recall the notation \(\boldsymbol{u}=(u,v,w)^{\rm T},\, \tilde{\nabla} = (\partial_x,\partial_y,\partial_{\zeta})^{\rm T}\) and \(\tilde{\nabla}_x=(\partial_x,\partial_y,0)^{\rm T}\). In addition \(J_{\Sigma}\) stand for the Jacobian of the mapping \(\Sigma\) as introduced in the sections~\ref{subsec:mapped_ref_sys} and defines the operator
\begin{equation}
	\label{appx:eq:mapping_Jacobian}
	\left((J_{\Sigma})^{\rm T}\tilde{D}\right)_x = \begin{pmatrix}
		\partial_x(h\bullet)-\partial_{\zeta}(\partial_x(\zeta h + h_b)\bullet)\\
		\partial_y(h\bullet)-\partial_{\zeta}(\partial_y(\zeta h + h_b)\bullet)\\
		\partial_{\zeta}\bullet
	\end{pmatrix}.
\end{equation}
To derive the moment system we also recall the standard approach
\begin{equation}
	\label{appx:eq:moment_approach_velo}
	\boldsymbol{u} = \boldsymbol{u}_m + \boldsymbol{u}_d = \boldsymbol{u}_m(t,x,y) + \sum_{j=1}^{N}\boldsymbol{\Lambda}_j(t,x,y)\phi_j(\zeta),
\end{equation}
with \(\boldsymbol{\Lambda}_j = (\alpha_j,\beta_j,\gamma_j)^T\). In addition we have for the pressure deviation
\begin{equation}
	\label{appx:eq:moment_approach_pressure}
	q = q_m + q_d = q_m(t,x,y) + \sum_{j=1}^{N}\kappa_j(t,x,y) \phi_j(\zeta),
\end{equation}
with \(\phi_i,\,i=0,\dots,N\) scaled Legendre polynomials on \(\left[0,1\right]\) that are normalized to \(\phi_i(0)=1\). With these definitions we are now able to derive the dispersive shallow moment system. The height balance does not change under Galerkin projection. So we consider the mapped divergence constraint first. This constraint is not present in the case of hydrostatic pressure. Using our introduced notation we calculate for the first terms of the constraint 
\begin{align*}
	\int_{0}^{1}\tilde{\nabla}_x\cdot(h\boldsymbol{u})\phi_i\,\mathrm{d}\zeta 
	&= \tilde{\nabla}_x\cdot(h\boldsymbol{u}_m)\int_{0}^{1}\phi_i\,\mathrm{d}\zeta +\sum_{j=1}^{N}\tilde{\nabla}_x\cdot\left(h\boldsymbol{\Lambda}_j\right)\int_{0}^{1}\phi_j\phi_i\,\mathrm{d}\zeta\\
	&=\tilde{\nabla}_{x}\cdot(h\boldsymbol{u}_m)\delta_{i0} + \sum_{j=1}^N\frac{\delta_{ij}\tilde{\nabla}_x\cdot(h\boldsymbol{\Lambda}_j)}{2i+1}.
\end{align*}
For the \(\zeta\)-derivative of \(w\) we have to use the boundary condition for \(w\) at the bottom, reading
\begin{equation}
	\label{appx:eq:boundary_bottom}
	w_{|_{\zeta =0}} = w_m+\sum_{j=1}^{N}\gamma_j = \boldsymbol{u}_{|_{\zeta=0}}\cdot\tilde{\nabla}_x h_b = \boldsymbol{u}_m \cdot \tilde{\nabla}_x h_b + \sum_{j=1}^{N} \boldsymbol{\Lambda}_j\cdot\tilde{\nabla}_x h_b.
\end{equation}
We get
\begin{align*}
	\int_0^1\partial_{\zeta}w\phi_i\,\mathrm{d}\zeta &=  
	\sum_{j=1}^{N}\gamma_j\int_{0}^{1}\phi_j'\phi_i\,\mathrm{d}\zeta =\sum_{j=1}^N\left(\gamma_j\left[\phi_i\phi_j\right]_0^1-\int_0^1\phi_i' \phi_j\,\mathrm{d}\zeta\right)\\
	&= w_m-\boldsymbol{u_m}\cdot\tilde{\nabla}_xh_b-\sum_{j=1}^{N}\boldsymbol{\Lambda}_j\cdot\tilde{\nabla}_xh_b - \gamma_j\left((-1)^{i+j}-\frac{G_{ij}}{2i+1}\right).
\end{align*}
Now we have two integrals left. Both are of the same form and using our notation they can be written in a compact way as
\[
-\partial_{\zeta}(\partial_x(\zeta h + h_b)u)-\partial_{\zeta}(\partial_y(\zeta h +h_b)v)=-\partial_{\zeta}(\tilde{\nabla}_x(\zeta h + h_b )\cdot \boldsymbol{u}).
\]
This gives us
\begin{align*}
	&\int_{0}^{1}\partial_{\zeta}(\tilde{\nabla}_x(\zeta h + h_b)\cdot\boldsymbol{u})\phi_i\,\mathrm{d}\zeta = \int_{0}^{1}\partial_{\zeta}(\tilde{\nabla}_x(\zeta h + h_b)\cdot\left(\boldsymbol{u}_m+\sum_{j=1}^{N}\boldsymbol{\Lambda}_j\phi_j)\right)\phi_i\,\mathrm{d}\zeta\\
	&\quad=\boldsymbol{u}_m\cdot\tilde{\nabla}_xh\delta_{i0} + \sum_{j=1}^{N}\frac{\tilde{\nabla}_xh\cdot\boldsymbol{\Lambda}_j\delta_{ij}}{2i+1}
	+\sum_{j=1}^{N}\tilde{\nabla}_xh\cdot\boldsymbol{\Lambda}_j \int_{0}^{1}\zeta\phi_j'\phi_i\,\mathrm{d}\zeta + \sum_{j=1}^{N}\tilde{\nabla}_xh_b\cdot\boldsymbol{\Lambda}_j\int_{0}^{1}\phi_j'\phi_i\,\mathrm{d}\zeta.
\end{align*}
For the remaining integrals we want to push the derivative to the polynomial independent of the sum. To achieve that we perform integration by parts:
\begin{align*}
	\frac{1}{2j+1}H_{ji}=\int_{0}^{1}\zeta \phi_j'\phi_i\,\mathrm{d}\zeta &= \left[\phi_j\phi_i\zeta\right]_0^1 - \int_{0}^{1}\phi_j\phi_i\,\mathrm{d}\zeta - \int_{0}^{1}\zeta\phi_j\phi_i'\,\mathrm{d}\zeta\\
	&=(-1)^{i+j}-\frac{\delta_{ij}}{2i+1}-\frac{1}{2i+1}H_{ij},\\
	\frac{1}{2j+1}G_{ji}=\int_{0}^{1}\phi_j'\phi_i\,\mathrm{d}\zeta &= \left[\phi_j\phi_i\right]_0^1-\int_{0}^{1}\phi_j\phi_i'\,\mathrm{d}\zeta
	= (-1)^{i+j}-1 -\frac{1}{2i+1}G_{ij}.
\end{align*}
We now use this in our formula above to get
\begin{align*}
	&\int_{0}^{1}\partial_{\zeta}\left(\tilde{\nabla}_x(\zeta h + h_b)\cdot\boldsymbol{u}\right)\phi_i\,\mathrm{d}\zeta=\boldsymbol{u}_m\cdot\tilde{\nabla}_xh\delta_{i0} + \sum_{j=1}^{N}\frac{\tilde{\nabla}_xh\cdot\boldsymbol{\Lambda}_j\delta_{ij}}{2i+1}\\
	&+\sum_{j=1}^{N}\tilde{\nabla}_x h \cdot\boldsymbol{\Lambda}_j \left((-1)^{i+j}-\frac{\delta_{ij}}{2i+1}-\frac{1}{2i+1}H_{ij}\right)+ \sum_{j=1}^{N}\tilde{\nabla}_xh_b\cdot\boldsymbol{\Lambda}_j\left((-1)^{i+j}-1 -\frac{1}{2i+1}G_{ij}\right).
\end{align*}
We can now put everything together to have our final equation
\begin{align}
	&h\tilde{\nabla}_x\cdot\boldsymbol{u}_m\delta_{i0}+\sum_{j=1}^{N}\frac{h\tilde{\nabla}_x\cdot\boldsymbol{\Lambda}_j\delta_{ij}}{2i+1}+w_m-\boldsymbol{u}_m\cdot\tilde{\nabla}_xh_b-\sum_{j=1}^{N}\boldsymbol{\Lambda}_j\cdot\tilde{\nabla}_xh_b+\sum_{j=1}^{N}\gamma_j\left((-1)^{i+j}-\frac{G_{ij}}{2i+1}\right)\nonumber\\
	&-\sum_{j=1}^{N}\tilde{\nabla}_x h \cdot\boldsymbol{\Lambda}_j \frac{1}{2j+1}H_{ji}- \sum_{j=1}^{N}\tilde{\nabla}_xh_b\cdot\boldsymbol{\Lambda}_j\frac{1}{2j+1}G_{ji}=0.\label{appx:eq:constraint_general}
\end{align}
In the special case of \(i=0\) we have from this equation
\begin{align}
	h\tilde{\nabla}_x\cdot\boldsymbol{u}_m + w_m -\boldsymbol{u}_m\cdot\tilde{\nabla}_xh_b +\sum_{j=1}^{N}\gamma_j(-1)^j 
	-\sum_{j=1}^{N}\tilde{\nabla}_xh\cdot\boldsymbol{\Lambda}_j (-1)^j - \sum_{j=1}^{N}\tilde{\nabla}_xh_b\cdot\boldsymbol{\Lambda}_j(-1)^j =0.\label{appx:eq:constraint_zero}
\end{align}
After we have found the moment approximation of the constraint we continue with the moment equations. Like for the hydrostatic case we consider all terms one by one. The first three terms can be calculated like in the hydrostatic case, given in~\cite{KowalskiTorrilhon_shallowmomentapprox}. The only different is that the vectors \(\boldsymbol{u}_m,\boldsymbol{u},\boldsymbol{u}_d\) resp. \(\boldsymbol{\Lambda}_i\) now have three components instead of two. But this does not change the calculation. So, we have
\begin{align*}
	\int_{0}^{1}\partial_t(h\boldsymbol{u})\phi_i\,\mathrm{d}\zeta &=\begin{cases}
		\partial_t(h\boldsymbol{u}_m),\qquad &i=0,\\
		\frac{\partial_t(h\boldsymbol{\Lambda}_i)}{2i+1}, &i\geq 1
	\end{cases},\\[2mm]
	\int_{0}^{1}\partial_x(hu\boldsymbol{u})\phi_i\,\mathrm{d}\zeta &= \begin{cases}
		\partial_x\left(h\left(u_m\boldsymbol{u}_m +\sum_{j=1}^{N}\frac{\alpha_j\boldsymbol{\Lambda}_j}{2j+1}\right)\right),\quad &i=0,\\
		\partial_x\left(h\left(\frac{u_m\boldsymbol{\Lambda}_i}{2i+1}+\frac{\alpha_i\boldsymbol{u}_m}{2i+1}+\sum_{j,k=1}^{N}\alpha_j\boldsymbol{\Lambda}_kA_{ijk}\right)\right), &i\geq 1
	\end{cases},\\[2mm]
	\int_{0}^{1}\partial_y(hv\boldsymbol{u})\phi_i\,\mathrm{d}\zeta &= \begin{cases}
		\partial_y\left(h\left(v_m\boldsymbol{u}_m +\sum_{j=1}^{N}\frac{\beta_j\boldsymbol{\Lambda}_j}{2j+1}\right)\right),\quad &i=0,\\
		\partial_y\left(h\left(\frac{v_m\boldsymbol{\Lambda}_i}{2i+1}+\frac{\beta_i\boldsymbol{u}_m}{2i+1}+\sum_{j,k=1}^{N}\beta_j\boldsymbol{\Lambda}_kA_{ijk}\right)\right), &i\geq 1
	\end{cases}.
\end{align*}
The same holds for the parts of the hydrostatic pressure
\begin{align*}
	\int_{0}^{1}\frac{g}{2}\tilde{\nabla}_x\left(h^2\right)\phi_i\,\mathrm{d}\zeta &=\frac{g}{2}\tilde{\nabla}_x\left(h^2\right)\int_{0}^{1}\phi_i\,\mathrm{d}\zeta = \begin{cases}
		\frac{g}{2}\tilde{\nabla}_x\left(h^2\right),\qquad &i=0.\\
		0, &i\geq 1
	\end{cases},\\
	\int_{0}^{1}hg\tilde{\nabla}_x h_b\phi_i\,\mathrm{d}\zeta &= hg\tilde{\nabla}_xh_b\int_{0}^{1}\phi_i\,\mathrm{d}\zeta=\begin{cases}
		hg\tilde{\nabla}_xh_b,\qquad &i=0,\\
		0, &i\geq 1
	\end{cases}.
\end{align*}
The term involving the vertical coupling operator can also be computed the same way. Here we state the calculation again using the notation of this section
\begin{align*}
	\int_{0}^{1}\partial_{\zeta}(h\omega\boldsymbol{u})\phi_i\,\mathrm{d}\zeta&= \int_{0}^{1}\partial_{\zeta}(h\omega\boldsymbol{u}_m)\phi_i\,\mathrm{d}\zeta + \int_{0}^{1}\partial_{\zeta}(h\omega\boldsymbol{u}_d)\phi_i\,\mathrm{d}\zeta\\
	&= -\int_{0}^{1}\boldsymbol{u}_m\partial_{\zeta} \left(\tilde{\nabla}_x\cdot\left(h\int_{0}^{\zeta}\boldsymbol{u}_d\,\mathrm{d}\hat{\zeta}\right)\right)\phi_i\,\mathrm{d}\zeta
	+\left[h\omega\boldsymbol{u}_d\phi_i\right]_0^1 - \int_{0}^{1}h\omega \boldsymbol{u}_d\phi_i'\,\mathrm{d}\zeta\\
	&= -\frac{\boldsymbol{u}_m\sum_{j=1}^{N} \tilde{\nabla}_x\cdot(h\boldsymbol{\Lambda}_j)\delta_{ij}}{2i+1} + \sum_{j,k=1}^{N}\boldsymbol{\Lambda}_k\tilde{\nabla}_x\cdot(h\boldsymbol{\Lambda}_j)B_{ijk}.
\end{align*}
More interesting is the pressure deviation that is not present in the hydrostatic case. Here we compute with the definition of the differential operator
\begin{equation}
	\label{appx:eq:pressure_deviation_comp}
	\left((J_{\Sigma})^{\rm T}\tilde{D}\right)_x q = \begin{pmatrix}
		\partial_x(hq) - \partial_{\zeta}(\partial_x(\zeta h + h_b)q)\\
		\partial_y(hq) - \partial_{\zeta}(\partial_y(\zeta h + h_b)q)\\
		\partial_{\zeta} q
	\end{pmatrix}.
\end{equation}
that for the integral holds
\begin{align*}
	&\int_{0}^{1}\left((J_{\Sigma})^{\rm T}\tilde{D}\right)_x q\phi_i\,\mathrm{d}\zeta = \int_{0}^{1}\left((J_{\Sigma})^{\rm T}\tilde{D}\right)_x q_m \phi_i\,\mathrm{d}\zeta + \sum_{j=1}^{N}\int_{0}^{1}\left((J_{\Sigma})^{\rm T}\tilde{D}\right)_x (\kappa_j\phi_j)\phi_i\,\mathrm{d}\zeta\\
	&\quad= \tilde{\nabla}_xq_m h\int_{0}^{1}\phi_i\,\mathrm{d}\zeta
	+\sum_{j=1}^{N}\int_{0}^{1}\begin{pmatrix}
		\partial_x(h\kappa_j)\phi_j - \kappa_j(\partial_x(\zeta h + h_b)\partial_{\zeta}\phi_j + \phi_j\partial_xh)\\
		\partial_y(h\kappa_j)\phi_j - \kappa_j(\partial_y(\zeta h + h_b)\partial_{\zeta}\phi_j + \phi_j\partial_yh)\\
		\kappa_j\partial_{\zeta}\phi_j
	\end{pmatrix} \phi_i\,\mathrm{d}\zeta\\
	&\qquad = h\tilde{\nabla}_xq_m\delta_{i0}+\sum_{j=1}^{N}\frac{h\tilde{\nabla}_x\kappa_j\delta_{ij}}{2i+1} - \tilde{\nabla}_xh\sum_{j=1}^{N}\frac{\kappa_jH_{ji}}{2j+1}
	- \tilde{\nabla}_xh_b\sum_{j=1}^{N}\frac{\kappa_j G_{ji}}{2j+1} + \sum_{j=1}^{N}\kappa_j\frac{G_{ji}}{2j+1}\boldsymbol{e}_{\zeta}.
\end{align*}
This is valid for every \(i=0,\dots,N\). In particular for \(i=0\) we get
\begin{align*}
	\int_{0}^{1}\left((J_{\Sigma})^{\rm T}\tilde{D}\right)_x q\,\mathrm{d}\zeta = \tilde{\nabla}_x(hq_m) + \tilde{\nabla}_xh_b\left(q_m + \sum_{j=1}^{N}\kappa_j\right) - \left(q_m + \sum_{j=1}^{N}\kappa_j\right)\boldsymbol{e}_{\zeta},
\end{align*}
where we used the boundary condition of the pressure deviation at the surface in the form
\[
q_{|_{\zeta = 1}}= 0 \quad \Leftrightarrow\quad q_m+\sum_{j=1}^{N}\kappa_j\phi_j(1)=0\quad \Leftrightarrow\quad q_m = - \sum_{j=1}^{N}(-1)^j\kappa_j.
\]
After these calculations we are now able to give the full nonlinear dispersive moment system. If we set everything together we end up with
\begin{align*}
	&\partial_t\left(h\boldsymbol{u}_m\right)+\partial_x\left(h\left(u_m\boldsymbol{u}_m+\sum_{j=1}^{N}\frac{\alpha_j\boldsymbol{\Lambda}_j}{2j+1}\right)\right)+\partial_y\left(h\left(v_m\boldsymbol{u}_m + \sum_{j=1}^{N}\frac{\beta_j\boldsymbol{\Lambda}_j}{2j+1}\right)\right)\\
	&+\frac{g}{2}\tilde{\nabla}_{x}(h^2)+hg\tilde{\nabla}_{x}h_b +\tilde{\nabla}_{x}(hq_m) + \tilde{\nabla}_{x}h_b\left(q_m + \sum_{j=1}^{N}\kappa_j\right) - \left(q_m + \sum_{j=1}^{N}\kappa_j\right)\boldsymbol{e}_{\zeta} =0,
\end{align*}
for \(i=0\) together with the constraint~\eqref{appx:eq:constraint_zero} and for the higher order moments \(i\geq 1\)
\begin{align*}
	&\frac{\partial_t\left(h\boldsymbol{\Lambda}_i\right)}{2i+1}+\partial_x\left(h\left(\frac{u_m\boldsymbol{\Lambda}_i}{2i+1}+\frac{\alpha_i\boldsymbol{u}_m}{2i+1}+\sum_{j,k=1}^{N}\alpha_j\boldsymbol{\Lambda}_k\frac{A_{ijk}}{2i+1}\right)\right)\\
	&+ \partial_y\left(h\left(\frac{v_m\boldsymbol{\Lambda}_i}{2i+1}+\frac{\beta_i\boldsymbol{u}_m}{2i+1}+\sum_{j,k=1}^{N}\beta_j\boldsymbol{\Lambda}_k\frac{A_{ijk}}{2i+1}\right)\right)
	-\frac{\boldsymbol{u}_m\tilde{\nabla}_x\cdot(h\boldsymbol{\Lambda}_i)}{2i+1}+\sum_{j,k=1}^{N}\boldsymbol{\Lambda}_k\tilde{\nabla}_x\cdot(h\boldsymbol{\Lambda}_j)\frac{B_{ijk}}{2i+1}\\
	&+\frac{h\tilde{\nabla}_{x}\kappa_i}{2i+1}-\tilde{\nabla}_{x}h\sum_{j=1}^{N}\frac{\kappa_jH_{ji}}{2j+1}
	-\tilde{\nabla}_{x}h_b\sum_{j=1}^{N}\frac{\kappa_jG_{ji}}{2j+1}+\sum_{j=1}^{N}\kappa_j\frac{G_{ji}}{2j+1}\boldsymbol{e}_{\zeta}=0.
\end{align*}
with the constraint~\eqref{appx:eq:constraint_general}.
This system is used in section~\ref{subsec:DSM} and for the numerical results in section~\ref{sec:Num_Experiments}. Some examples for specific \(N\) are given in section~\ref{sec:examples_moment_systems} and in the appendix~\ref{appx:subsec:higher_order_examples}.

\subsection{Pressure constraint for DSM0}\label{appx:subsec:pressure_constraint_dsm0}

Here, we give a schematic calculation for the zeroth level system, which is in dimensionless form given by the equations~\eqref{eq:dsm0} with constraint~\eqref{eq:dsm0_constr}. To get the reformulated constraint~\eqref{eq:dsm0_poisson} we perform the steps from section~\ref{sec:reformulation_nonlinear_constraint}:
\begin{enumerate}
	\item Solve the evolution equations for height and velocity for the time derivatives of the corresponding variables and apply the height balance
	\begin{equation}
		\label{eq:dsm0_solved_time_height}
		\partial_th = -\partial_x(hu_m)
	\end{equation}to replace the time derivatives in the evolution equations
	\begin{align}
		\label{eq:dsm0_solved_time_hori_velo}
		\partial_tu_m &= -\frac{1}{h\Fr^2}\left( -u_m\partial_x(hu_m) + \partial_x\left(\frac{h^2}{2} + hq_m + \Fr^2hu_m^2\right)+ \frac{\Fr^2}{\He}\partial_xh_b(h+q_m)\right),\\
		\label{eq:dsm0_solved_time_vert_velo}
		\partial_tw_m &= -\frac{1}{h}\left(-w_m\partial_x(hu_m) + \partial_x(hu_mw_m) - \frac{1}{\He^2\Fr^2}q_m \right).
	\end{align}
	\item Take the time derivative of the constraint~\eqref{eq:dsm0_constr}
	\begin{equation}
		\label{eq:dsm0_constr_time_deri}
		\partial_th\partial_xu_m + h\partial_x\partial_tu_m = -\partial_tw_m+\frac{1}{\He}\partial_tu_m\partial_xh_b.
	\end{equation}
	\item Take the \(x\)-derivative of the constraint~\eqref{eq:dsm0_constr}
	\begin{equation}
		\label{eq:dsm0_constr_x_deri}
		u_mh\partial_xh\partial_xu_m + u_mh^2\partial_{xx}u_m +u_mh\partial_xw_m -\frac{1}{\He}u_mh\partial_xu_m\partial_xh_b - \frac{1}{\He}u_m^2h\partial_{xx}h_b = 0.
	\end{equation}
	\item Adding \(h\,\cdot\)~\eqref{eq:dsm0_constr_time_deri} and \(hu_m\cdot\)~\eqref{eq:dsm0_constr_x_deri} and replacing all time derivatives with the help of~\eqref{eq:dsm0_solved_time_height}, \eqref{eq:dsm0_solved_time_hori_velo} and~\eqref{eq:dsm0_solved_time_vert_velo}
	\begin{align*}
		&-h\partial_x(hu_m)\partial_xu_m + h^2\partial_x\eqref{eq:dsm0_solved_time_hori_velo} + h\eqref{eq:dsm0_solved_time_vert_velo} -\frac{1}{\He}h\eqref{eq:dsm0_solved_time_hori_velo}\partial_xh_b + h^2u_m^2 \partial_xh\partial_xu_m + h^3u_m^2 \partial_{xx}u_m \\
		&+h^2u_m^2\partial_xw_m - \frac{1}{\He}h^2u_m^2\partial_xu_m\partial_xh_b -\frac{1}{\He}h^2u_m^3\partial_{xx}h_b = 0.
	\end{align*}
	This gives after simplification the final reformulated constraint of the nonlinear zeroth level system
	\begin{align*}
		h^2 \partial_{xx}q_m=&-\left(hq_m+h^2\right)\partial_{xx}h- \left(\frac{1}{\He} hq_m+\frac{\Fr^2}{\He} hu_m^2+\frac{1}{\He} h^2\right)\partial_{xx} h_b\\&+
		\frac{1}{\He} 2 q_m \dx h_b \partial_x h
		+\frac{1}{\He} h \dx h_b \partial_x h 
		-h \partial_x h  \partial_x q_m
		+q_m (\partial_x h)^2 \\
		&+\frac{1}{\He^2} h (\partial_x h_b)^2 
		-2 \Fr^2 h^2 (\partial_x u_m)^2
		+\frac{1}{\He^2} q_m(\dx h_b)^2+\frac{1}{\He^2} q_m.
	\end{align*}
\end{enumerate}


\printbibliography

\end{document}